\documentclass[useAMS,usenatbib,usegraphicx]{mn2e}

\title[Supermassive black hole mass functions at intermediate
redshifts]{Supermassive Black Hole Mass Functions at Intermediate
Redshifts from Spheroid and AGN Luminosity Functions}

\author[N. Tamura, K. Ohta, \& Y. Ueda]{Naoyuki
Tamura$^{1}$\thanks{E-mail:naoyuki.tamura@durham.ac.uk}, Kouji
Ohta$^{2}$, \& Yoshihiro Ueda$^{2, 3}$\\ $^{1}$Department of Physics,
University of Durham, South Road, Durham, DH1 3LE, UK\\ $^{2}$Department
of Astronomy, Kyoto University, Kyoto 606-8502, Japan\\ $^{3}$Institute
of Space and Astronautical Sciences, 3-1-1 Yoshinodai, Sagamihara-shi,
Kanagawa 229-8510, Japan}

\begin{document}

\date{}

\pagerange{\pageref{firstpage}--\pageref{lastpage}} \pubyear{2005}

\maketitle

\label{firstpage}

\begin{abstract}

Redshift evolution of supermassive black hole mass functions (BHMFs) is
investigated up to $z \sim 1$. BHMFs at intermediate redshifts are
calculated in two ways. One way is from early-type galaxy luminosity
functions (LFs); we assume an $M_{\rm BH} - L_{\rm sph}$ correlation at
a redshift by considering a passive evolution of $L_{\rm sph}$ in the
local relationship. The resultant BHMFs (spheroid-BHMFs) from LFs of red
sequence galaxies indicates a slight decrease of number density with
increasing redshift at $M_{\rm BH} \geq 10^{7.5-8} M_{\odot}$. Since a
redshift evolution in slope and zeropoint of the $M_{\rm BH} - L_{\rm
sph}$ relation is unlikely to be capable of making such an evolution in
BHMF, the evolution of the spheroid-BHMFs is perhaps due mainly to the
decreasing normalization in the galaxy LFs. We also derive BHMFs from
LFs of morphologically selected early-type galaxies. The resultant BHMFs
are similar to those from the red sequence galaxies, but show a small
discrepancy at $z \sim 1$ corresponding to an increase of SMBH number
density by $\sim$ 0.3 dex. We also investigate how spheroid-BHMFs are
affected by uncertainties existing in the derivation in detail.

The other way of deriving a BHMF is based on the continuity equation for
number density of SMBHs and LFs of active galactic nucleus (AGN). The
resultant BHMFs (AGN-BHMFs) show no clear evolution out to $z = 1$ at
$M_{\rm BH} \geq 10^8 M_{\odot}$, but exhibit a significant decrease
with redshift in the lower mass range. Interestingly, these AGN-BHMFs
are quite different in the range of $M_{\rm BH} \leq 10^8 M_{\odot}$
from those derived by Merloni (2004), where the fundamental plane of
black hole activity is exploited.

Comparison of the spheroid-BHMFs with the AGN-BHMFs suggests that at
$M_{\rm BH} \geq 10^{8} M_{\odot}$, the spheroid-BHMFs are broadly
consistent with the AGN-BHMFs out to $z \sim 1$. Although the decrease
of SMBH number density with redshift suggested by the spheroid-BHMFs is
slightly faster than that suggested by the AGN-BHMFs, we presume this to
be due at least partly to a selection effect on the LFs of red sequence
galaxies; the colour selection could miss spheroids with blue colours.
The agreement between the spheroid-BHMFs and the AGN-BHMFs appears to
support that most of the SMBHs are already hosted by massive spheroids
at $z\sim 1$ and they evolve without significant mass growth since then.

\end{abstract}


\section{INTRODUCTION}

Recent observations provide evidence that a mass of a supermassive black
hole (SMBH) in a galactic nucleus is tightly correlated with a mass or
luminosity of a spheroid component of its host galaxy (e.g., Magorrian
et al. 1998; Marconi \& Hunt 2003, MH03 hereafter). The tight
correlation suggests the presence of strong evolutionary link between
SMBH and spheroid component. Using the relation, an SMBH mass function
(BHMF) can be derived from local galaxy luminosity function (LF) or
velocity dispersion function. Meanwhile, a local BHMF can also be
calculated from cosmological evolution of LFs of active galactic nuclei
(AGNs) by integrating the continuity equation for number density of
SMBHs, where mass accretion onto a SMBH is assumed to power an AGN and
grow the central SMBH (e.g., Cavaliere, Morrison \& Wood 1971; Small \&
Blandford 1992; Marconi et al. 2004; Shankar et al. 2004). BHMFs derived
by this method can now be more reliable than before thanks to updated
LFs of hard X-ray selected AGNs (Ueda et al.  2003), which is more
complete to obscured AGNs. Marconi et al. (2004) and Shankar et
al. (2004) demonstrate that the local densities of SMBHs and BHMFs
derived with these two ways agree with each other, if one adopts
reasonable values for accretion efficiency and Eddington ratio.
Furthermore, Marconi et al. (2004) indicate that the cosmic history of
mass accretion rate density delineates that of star-formation rate
density, and the ratio of the latter to the former is about 4000, which
agrees with the mass ratio of a spheroid to a central SMBH, again
strongly suggesting the co-evolution of SMBHs and spheroids.

In studying the co-evolution of SMBHs and host spheroids in further
detail, one approach is to investigate redshift evolution of BHMF and
correlation between black hole mass ($M_{\rm BH}$) and spheroid
luminosity ($L_{\rm sph}$) or mass. While $M_{\rm BH}$ has been measured
for a substantial number of high redshift QSOs (e.g., Shields et
al. 2003; McLure \& Dunlop 2004), it is technically challenging to
directly measure dormant SMBHs at cosmological distances. However, BHMFs
at high redshifts can be computed using AGN LFs and the continuity
equation in the same way as in the local universe. Also, we are now in a
reasonable position to be able to study BHMFs at intermediate redshifts
by using LFs of early-type galaxies. Early-type galaxy LFs have recently
been derived out to $z \sim 1$ with good statistics from intensive
imaging surveys such as COMBO-17 (Wolf et al. 2003; Bell et al. 2004b).
Consequently, one can compare BHMFs from the galaxy LFs with those from
AGN LFs, which may provide clues to understand the co-evolution of AGNs
and host spheroids.

In this paper, we will derive BHMFs at intermediate redshifts from
galaxy LFs and AGN LFs and investigate the redshift evolutions. When
converting galaxy LFs to BHMFs, a correlation between $M_{\rm BH}$ and
$L_{\rm sph}$ is utilized. Although a correlation between $M_{\rm BH}$
and bulge effective stellar velocity dispersion ($\sigma_e$) is claimed
to be tighter than that between $M_{\rm BH}$ and $L_{\rm sph}$
(Ferrarese \& Merritt 2000; Gebhardt et al. 2000), measuring $\sigma_e$
at cosmological distance is very hard and we thus adopt the $M_{\rm BH}
- L_{\rm sph}$ relation in this work. It should also be emphasized that
MH03 show the $M_{\rm BH} - L_{\rm sph}$ relation is as tight as that
between $M_{\rm BH}$ and $\sigma_e$ if only SMBHs whose masses are
securely determined are used in the analysis. In addition, MH03 suggest
that the intrinsic scatter of the relation in $B$-band is as small as in
NIR bands. One should be aware that the derivation of a BHMF from galaxy
LFs contains several processes which may give significant uncertainties
to resultant BHMFs. Quantifying uncertainties in BHMFs when calculated
from galaxy LFs is also aimed at in this paper.

The layout of this paper is as follows. In the next section
(\S~\ref{sphbhmf}), we describe the procedure to derive a BHMF from an
early-type galaxy LF and calculate BHMFs from early-type galaxy LFs up
to $z \sim 1$. We also investigate how BHMFs are affected by
uncertainties and unconstrained parameters existing in this derivation.
In \S~\ref{agnbhmf}, we derive BHMFs from AGN LFs and examine its
redshift evolution. In \S~\ref{discussion}, we compare the BHMFs from
early-type galaxy LFs with those from AGN LFs and discuss the results.
Throughout this paper, we adopt the cosmological model with $H_0 = 70$
km s$^{-1}$ Mpc$^{-1}$, $\Omega_M = 0.3$ and $\Omega_{\Lambda} = 0.7$
unless otherwise stated.

\section{BHMF FROM EARLY-TYPE GALAXY LF}\label{sphbhmf}

\subsection{Derivation of BHMF}

In this paper, we mainly use the LFs obtained by Bell et al. (2004b)
from the COMBO-17 survey. The large survey area with the moderate depth
brings a large number of galaxies, and the multi-band photometry
covering from 3640 \AA~ to 9140 \AA~with the 17 broad- and medium-band
filters allows one to accurately determine the photometric redshifts.
Both of these aspects are important to derive luminosity functions with
good accuracy and the large survey area is especially useful to estimate
its cosmic variance. They firstly investigated the rest-frame $U-V$
vs. $M_V$ colour-magnitude diagram of galaxies at a certain redshift and
found the red sequence consistent with the colour-magnitude relation
well established for the early-type galaxy population. They select
galaxies on the red sequence and derive their rest-frame $B$-band LFs at
redshifts from 0.25 to 1.05.  Bell et al. (2004b) also derived the
$B$-band LF of the local red sequence galaxies using the SDSS EDR data
(Stoughton et al. 2002) by transforming the SDSS $ugr$ system to the
standard $UBV$ system. We use the Schechter functions fitted to these
LFs in the following analyses. A part of the survey area of the COMBO-17
was imaged with the HST/ACS and most of the galaxies on the red sequence
($\sim 85$ \%) indeed show early-type morphology (Bell et al. 2004a). It
should be noted that the colour selection based on the red sequence is
presumed to exclude blue ellipticals with star formation activity and/or
young stellar population. In addition, small bulges in late-type
galaxies can be missed, which perhaps results in a deficiency of the
light part of a BHMF.

In order to convert these LFs to BHMFs, firstly the total galaxy
luminosities need to be transformed to the spheroid luminosities using
bulge-to-total luminosity ratios ($B/Ts$). $B/T$s of early-type galaxies
at intermediate redshifts are, however, not well constrained
observationally. Im et al. (2002) selected morphologically early-type
galaxies at intermediate redshifts (most of the galaxies are at $z = 0.2
- 1.0$) from the HST/WFPC2 data for DEEP Groth-Strip Survey (DGSS) based
on $B/T$s ($> 0.4$), which are estimated by fitting radial surface
brightness profiles with $r^{1/4}$ spheroid and exponential disk.
According to their catalog, the morphologically selected early-type
galaxies have a $B/T$ of 0.7 on average, although there is a substantial
scatter. The $B/T$s do not depend on redshift or luminosity in their
sample. Therefore we adopt $B/T = 0.7$ in this study, independently of
redshift and galaxy luminosity.

Once LFs of spheroidal components are calculated from the early-type
galaxy LFs, they can be transformed to BHMFs using an $M_{\rm BH} -
L_{\rm sph}$ relation. In this study, we use those derived by MH03 for
their galaxies in Group 1, for which measurements of $M_{\rm BH}$ and
$L_{\rm sph}$ are considered to be reliable. The relation is derived for
$L_{\rm sph}$ in the $B$, $J$, $H$, and $K$ bands. Although the relation
in $K$-band is generally preferred because $K$-band luminosity is
considered to be the most reliable indicator of the stellar mass of a
galaxy, MH03 show the intrinsic scatter of the relation is almost
independent of the bands for the galaxies in Group 1. Since we will
mainly use the $B$-band LFs by Bell et al. (2004b), we adopt the
$B$-band relation described as follows:

\begin{equation}
 \log M_{\rm BH}=(1.19 \pm 0.12)(\log L_{\rm sph}-10.0)+(8.18 \pm 0.08). \label{eq:bs}
\end{equation}

\noindent
In applying the $M_{\rm BH} - L_{\rm sph}$ relation to an LF, the
intrinsic scatter of the relation needs to be considered; we adopt
$\Delta \log M_{\rm BH} = 0.32$ according to MH03.

In calculating a BHMF from a spheroid LF at a high redshift, an
evolution of the $M_{\rm BH} - L_{\rm sph}$ relation needs to be
considered. In this study, we basically consider only the effect of
passive evolution in $L_{\rm sph}$. It should be noted that if all
spheroid components evolve only passively (with no growth of SMBHs),
then the BHMF at the intermediate redshift does not change from that at
$z = 0$. In fact, it is not obvious that all spheroids evolve passively
at intermediate redshifts. Any other evolutions in galaxy LFs and the
$M_{\rm BH} - L_{\rm sph}$ relation would cause the disagreement between
BHMFs from galaxy LF and AGN LF, which will be discussed in
\S~\ref{discussion}. We describe the passive luminosity evolution as
$M_B(z) = M_B(z=0) - Qz$ and assume $Q$ to be 1.4. This is evaluated
using PEGASE Ver 2.0 (Fioc \& Rocca-Volmerange 1997) for a stellar
population with the solar metallicity formed at $z = 4$ (the age at $z =
0$ is 12 Gyr) in a single starburst with an $e$-folding time of 1
Gyr. This $Q$ value is consistent with the evolution of characteristic
luminosity in the COMBO-17 LFs ($z \geq 0.25$).

\subsection{BHMFs from early-type galaxy LFs up to $z \sim 1$}

\begin{figure}
 \centering \includegraphics[width=7cm,keepaspectratio,clip]{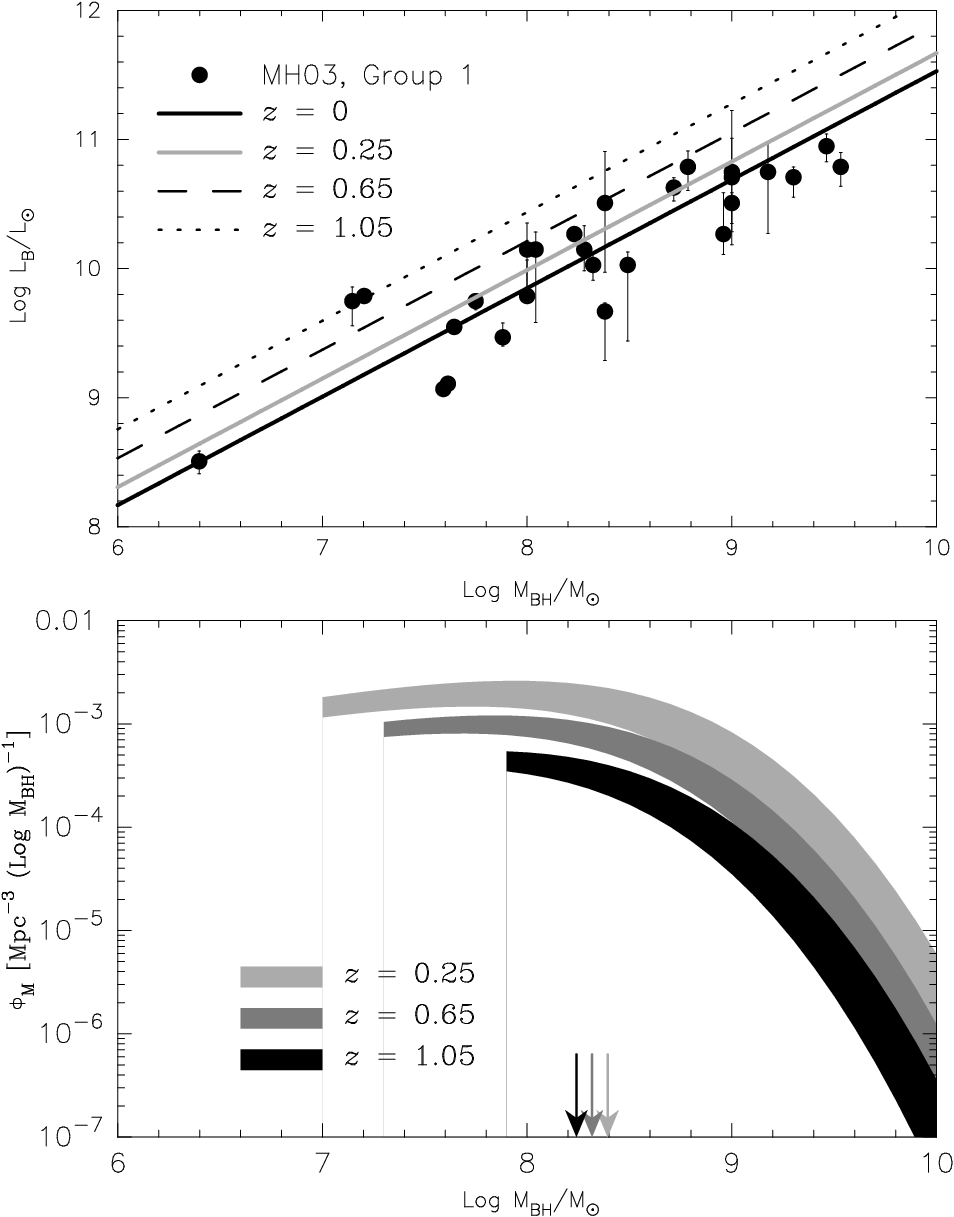}
 \caption{{\it Upper panel}: The correlation between $M_{\rm BH}$ and
 $B$-band spheroid luminosity. Black solid line indicates the relation
 at $z=0$ (equation (\ref{eq:bs})) fitted to the data points by MH03 for
 their galaxies in Group 1. Grey solid, black dashed, and dotted line is
 the relation expected at $z = 0.25, 0.65$ and 1.05, respectively, when
 a passive evolution of spheroid luminosity is considered. {\it Lower
 panel}: BHMFs transformed from the COMBO-17 LFs at redshifts of 0.25,
 0.65, and 1.05 are indicated with shaded regions, of which widths show
 the errors in $M_{B}^{\ast}$ and $\phi^{\ast}$ (Bell et al. 2004b). The
 lower mass cutoff of these BHMFs corresponds to the lowest luminosity
 of the data points in the original LF. The characteristic $M_{\rm BH}$
 corresponding to $M_{B}^{\ast}$ at each redshift is indicated by arrow
 with the same colour as of the BHMF.}  \label{relevo}
\end{figure}

Figure \ref{relevo} shows the BHMFs calculated using the COMBO-17 LFs
with the prescription described above. In the upper panel, the $B$-band
$L_{\rm sph}$ of the galaxies in Group 1 by MH03 is plotted against
$M_{\rm BH}$. Black solid line indicates the best-fitting regression
line to the data (i.e., the relation at $z = 0$). Grey solid, black
dashed, and black dot-dashed line is the relation expected at $z = 0.25,
0.65$ and 1.05, respectively, when the passive luminosity evolution is
considered.  In the lower panel, BHMFs transformed from the COMBO-17 LFs
in the rest-frame $B$ band at these redshifts are indicated with shaded
regions. The envelope of each BHMF shows the errors of $M_{B}^{\ast}$
and $\phi^{\ast}$ in the Schechter function fit to the data (Bell et al.
2004b); the upper and lower bound is defined by a BHMF with the largest
and smallest $L_{B}^{\ast}$ and $\phi^{\ast}$ within the fitting error,
respectively. The uncertainty in $\phi^{\ast}$ is dominated by cosmic
variance (Bell et al. 2004b) and we adopt the larger value of the two
different estimates by Bell et al. (2004b). Each BHMF is indicated down
to the black hole mass corresponding to the lowest luminosity among the
data points in the LF. The characteristic $M_{\rm BH}$ corresponding to
$M_{B}^{\ast}$ at each redshift is indicated by arrow with the same
colour as the BHMF.

The BHMFs in Figure \ref{relevo} exhibit a redshift evolution.  Since
the characteristic $M_{\rm BH}$ does not largely change with redshift,
this evolution is perhaps due to a density evolution of BHMFs
corresponding to the decreasing normalization of the red sequence galaxy
LFs (Bell et al. 2004b). We will discuss other possibilities in
\S~\ref{evorel}.

Currently, the COMBO-17 survey provides one of the largest and most
useful databases to study galaxies out to $z \sim 1$ and the derived
early-type galaxy LFs are therefore considered to be the most reliable
so far. Nevertheless it is worth investigating the evolution of BHMF
using LFs determined with other data sets, in particular to see whether
the result is sensitive to selection criterion for early-type
galaxy. Since early-type galaxies are selected from the colour-magnitude
diagram in COMBO-17, we investigate the BHMFs derived from LFs of
morphologically selected early-type galaxies by Im et al. (2002; the
HST/WFPC2 data for DGSS are used) and by Cross et al. (2004; the data
were taken with the HST/ACS in the guaranteed time observations). In
these studies, early-type galaxies are selected based on the analyses of
radial surface brightness profiles. In Im et al. (2002), the rest-frame
$B$-band LFs are derived in the two redshift bins: $0.05 < z < 0.6$ and
$0.6 < z < 1.2$, while in Cross et al. (2004), they are obtained at $0.5
< z < 0.75$ and $0.75 < z < 1.0$.

\begin{figure}
 \centering
 \includegraphics[width=7cm,keepaspectratio,clip]{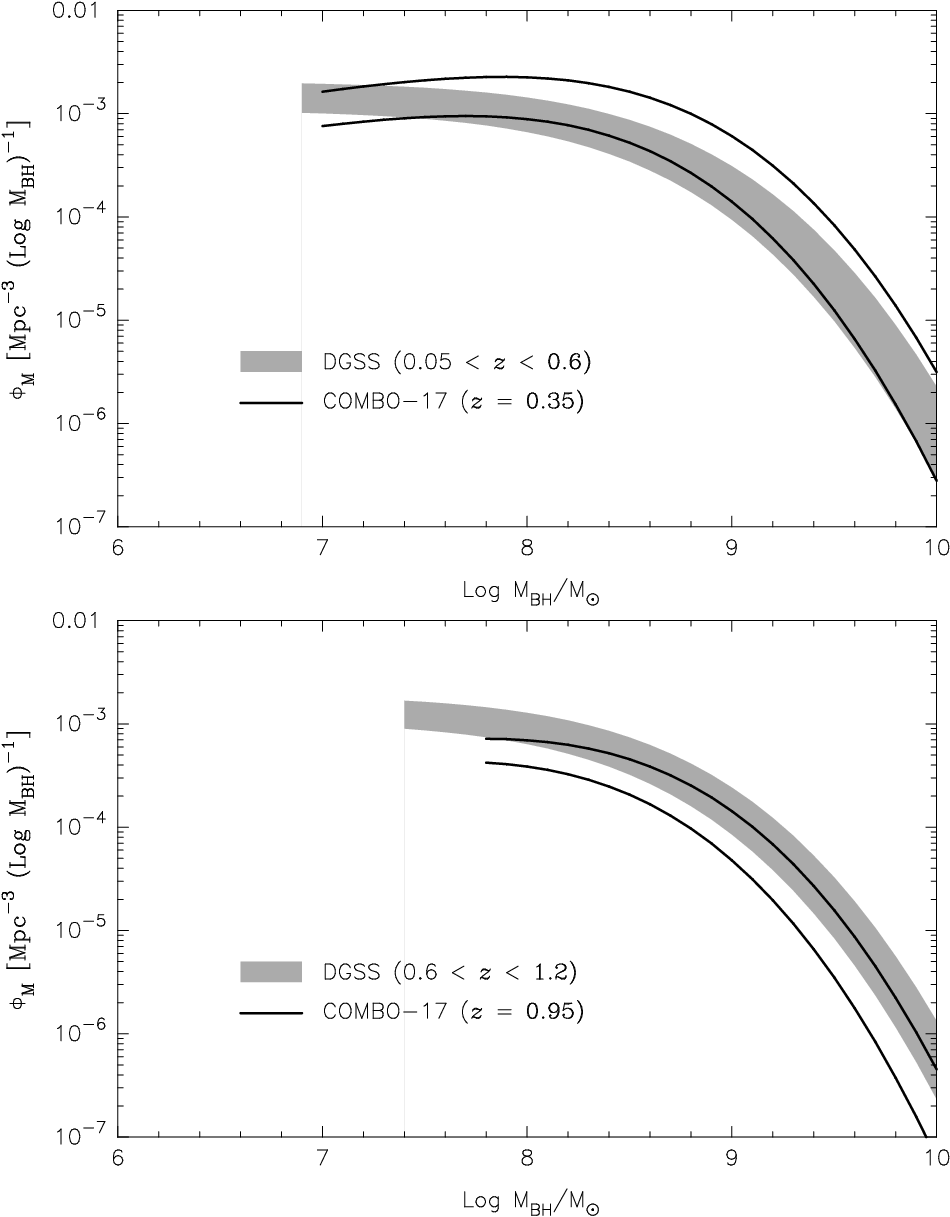}
 \caption{The BHMFs converted from the early-type galaxy LFs in the DGSS
 (Im et al. 2002) are indicated with shaded regions, whose widths
 represent the uncertainties of the BHMFs due to the fitting errors of
 $M_{B}^{\ast}$ and $\phi^{*}$ in the LFs. The pair of solid lines
 describes the BHMF from COMBO-17 at similar redshifts and the
 separation of the two lines indicates the uncertainty of the BHMF. The
 low mass cutoff of the BHMF corresponds to the lowest luminosity among
 the data points in the LF.} \label{im}
\end{figure}

\begin{figure}
 \centering
 \includegraphics[width=7cm,keepaspectratio,clip]{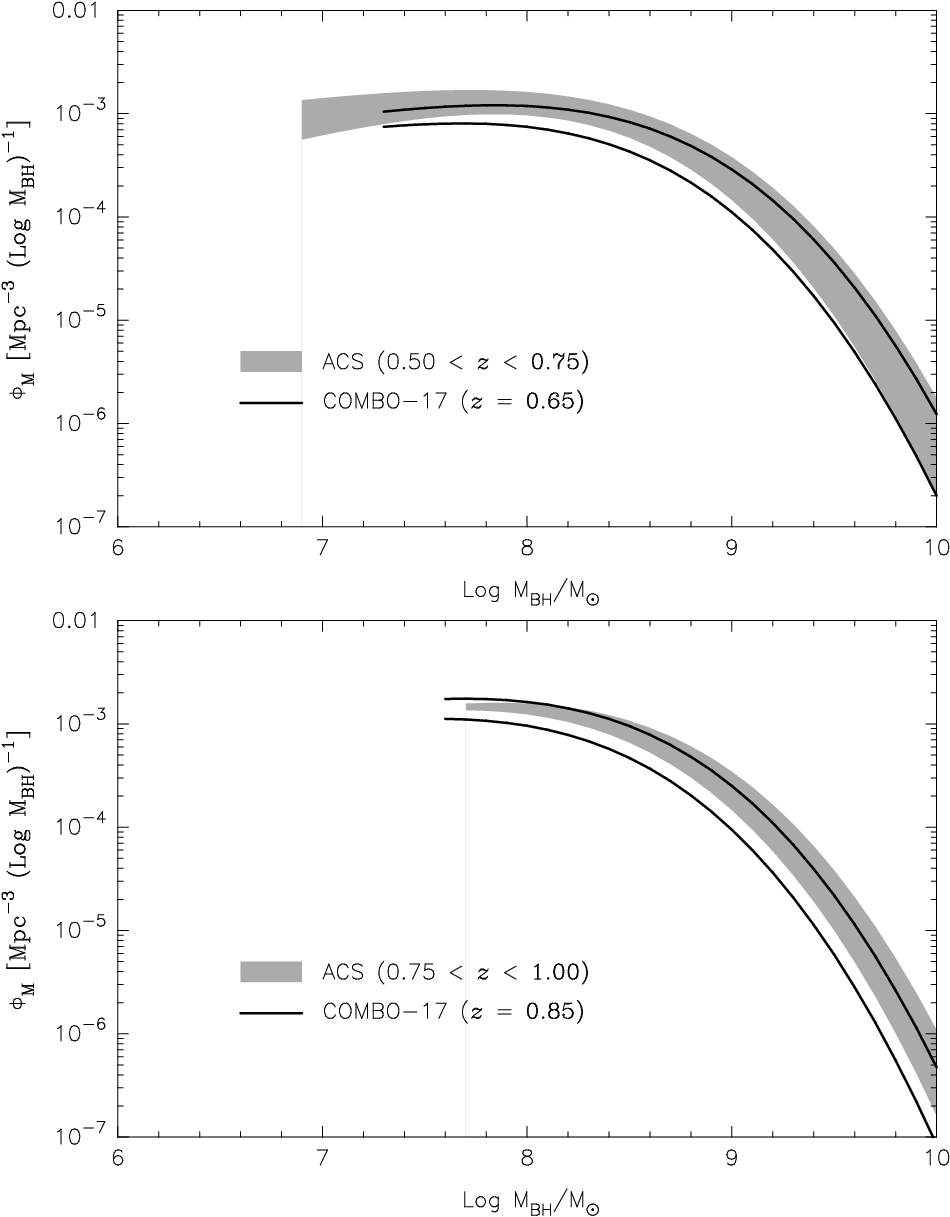}
 \caption{Same as Figure \ref{im}, but for the BHMFs converted from the
 early-type galaxy LFs computed by Cross et al. (2004) using the HST/ACS
 data.} \label{cross}
\end{figure}

In Figures \ref{im} and \ref{cross}, the BHMFs converted from the LFs of
morphologically selected early-type galaxies by Im et al. (2002) and
Cross et al. (2004), respectively, are presented by shaded regions,
showing the uncertainties due to the fitting errors of $M_{B}^{\ast}$
and $\phi^{*}$ in the LFs. Again these BHMFs are calculated by assuming
$B/T = 0.7$ and using the $B$-band $M_{\rm BH} - L_{\rm sph}$ relation
by MH03 corrected for passive luminosity evolutions. The low mass cutoff
of the BHMF corresponds to the lowest luminosity among the data points
in the original LF. The BHMF from the COMBO-17 LF at a similar redshift
is also indicated with solid lines, showing the upper and lower bounds
defined by considering the errors of $M_{B}^{\ast}$ and $\phi^{\ast}$ in
the LF. These comparisons demonstrate that the BHMFs from the
morphologically selected early-type galaxy LFs are consistent with those
from the COMBO-17 survey. It should be mentioned that the BHMFs obtained
from COMBO-17 LFs tend to lie below those from the morphologically
selected early-type galaxy LFs at the higher redshifts; the discrepancy
in number density of SMBHs at a given $M_{\rm BH}$ is estimated to be
$\sim$ 0.3 dex. This may indicate that the contribution of blue
spheroids which are not included in the COMBO-17 LFs becomes larger
towards $z \sim 1$.

\subsection{Source of uncertainty in the derivation of a BHMF}
\label{uncertainty}

In the previous subsections, we showed the procedure to convert galaxy
LFs to BHMFs at intermediate redshifts and presented the resultant BHMFs
up to $z \sim 1$. However, there are several sources of uncertainty in
the derivation.  Some of them are related to the $M_{\rm BH} - L_{\rm
sph}$ relation such as fitting error of the relation to the data points,
uncertainty of its intrinsic scatter, and choice of $M_{\rm BH} - L_{\rm
sph}$ relations. Others are due to the fact that $B/T$ value and passive
luminosity evolution are not well constrained. In what follows, we will
investigate how BHMFs are affected by these uncertainties. Unless
otherwise noted, we will begin all the calculations to derive BHMFs with
the LF of red sequence galaxies at $z = 0$ (Bell et al. 2004b, the SDSS
EDR data are used). We call this input LF ``iLF'' hereafter.

\subsubsection{Fitting error and uncertainty of intrinsic scatter in
$M_{\rm BH} - L_{\rm sph}$ relation} \label{fitandscat}

\begin{figure}
 \centering
 \includegraphics[width=7cm,keepaspectratio,clip]{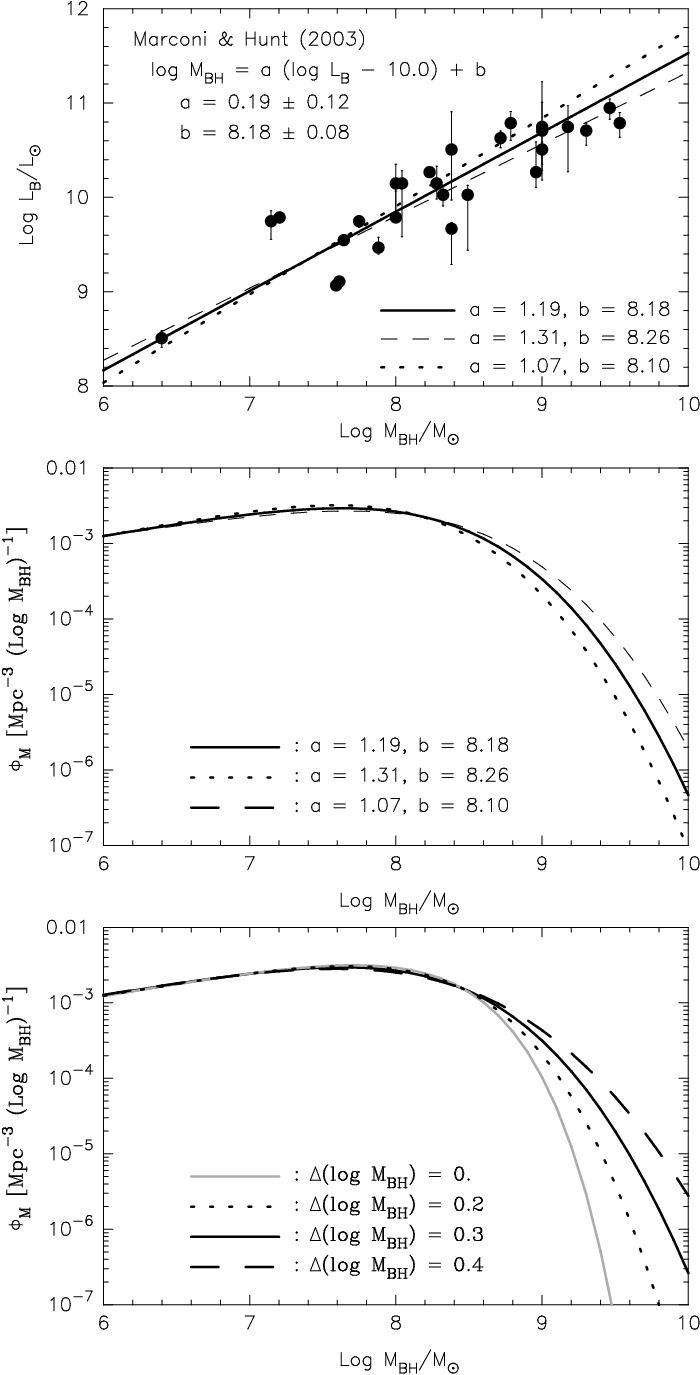}
 \caption{{\it Top panel}: The $M_{\rm BH}$s and $L_{\rm sph}$s of the
 galaxies in Group 1 by MH03 are plotted and the best-fitting regression
 line to the data points is indicated by solid line. Dashed (dotted)
 line shows the relation which results in a BHMF most (least) biased
 towards the massive end within the 1 $\sigma$ fitting errors,
 respectively, when applied to an LF. {\it Middle panel}: BHMFs are
 computed from iLF with the $M_{\rm BH} - L_{\rm sph}$ relations
 demonstrated above and they are indicated with the same line styles as
 in the top panel. The intrinsic scatter of the relation ($\Delta \log
 M_{\rm BH} = 0.32$; MH03) is considered to derive the BHMFs. {\it
 Bottom panel}: BHMFs are calculated for several values of intrinsic
 scatter of the relation; 0 (grey solid line), 0.2 (dotted line), 0.3
 (black solid line), and 0.4 (dashed line).} \label{fitting}
\end{figure}

In Figure \ref{fitting}, the fitting error in the $B$-band $M_{\rm BH} -
L_{\rm sph}$ relation estimated by MH03 is demonstrated in the top
panel. Three relations are shown here: One is the best-fitting
regression line (solid line), and the other two are those which result
in the most massive or least massive BHMFs within the $\pm 1 \sigma$
fitting error. The BHMF using either of the three relations is indicated
in the middle panel. In calculating the BHMFs, the intrinsic scatter
$\Delta \log M_{\rm BH} = 0.32$ (MH03) is taken into account. This shows
that the part of a BHMF at $M_{\rm BH} \geq 10^{8.5} M_{\odot}$ is
affected by this uncertainty.

We also calculate BHMFs for several values of intrinsic scatter of the
relation and show the results in the bottom panel; 0 (grey solid line),
0.2 (dotted line), 0.3 (black solid line), and 0.4 (dashed line). This
affects again the massive end of a BHMF ($M_{\rm BH} \geq 10^{8.5}
M_{\odot}$). It is suggested by MH03 to be between 0.3 and 0.4 (see also
McLure \& Dunlop 2002), but we note that if observational errors in
$M_{\rm BH}$ are underestimated, the intrinsic scatter would be smaller.

\subsubsection{$M_{\rm BH} - L_{\rm sph}$ relations in $B$ band and $K$
band} \label{colourcorr}

\begin{figure}
 \centering
 \includegraphics[width=7cm,keepaspectratio,clip]{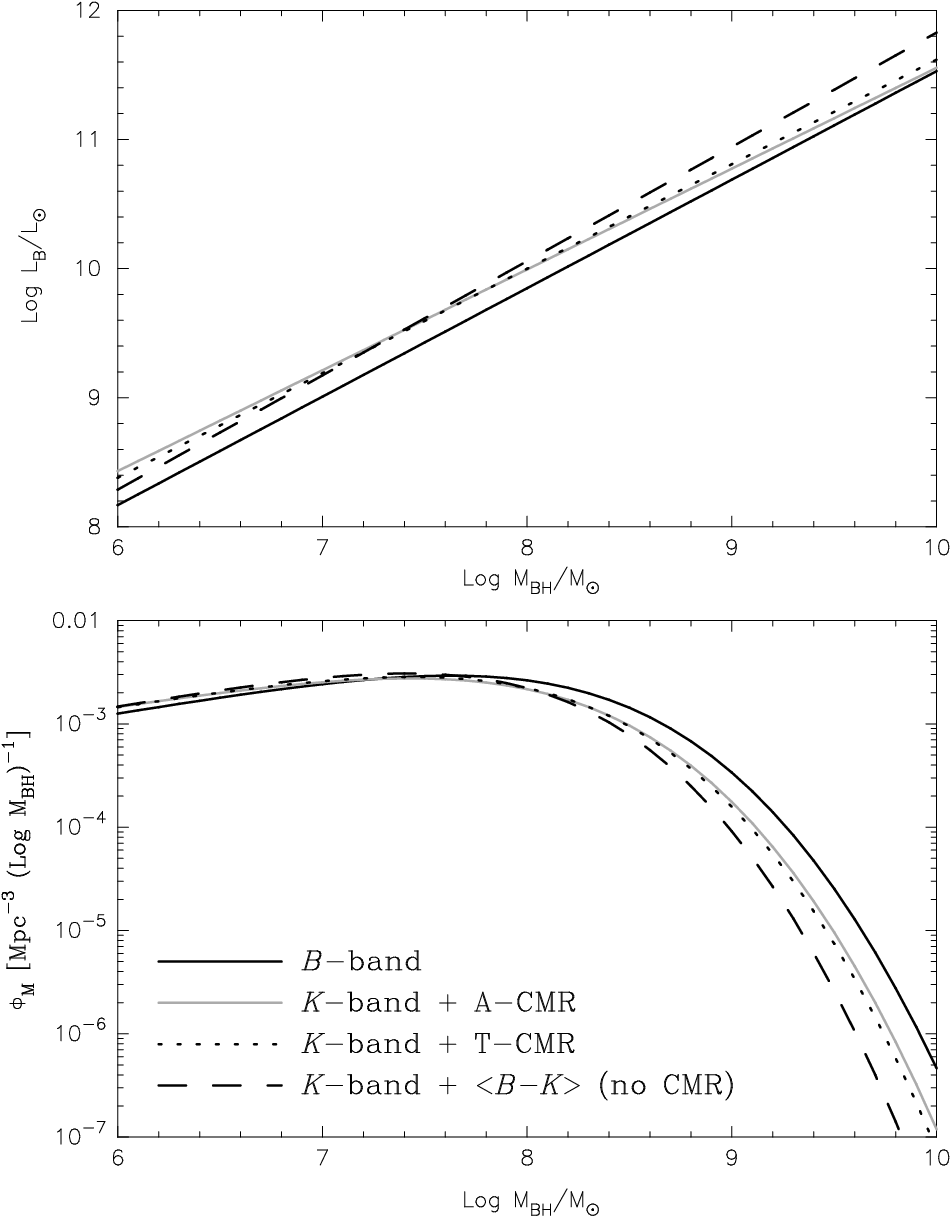}
 \caption{{\it Upper panel}: $B$-band $M_{\rm BH} - L_{\rm sph}$
 relations. Black solid line is the $B$-band $M_{\rm BH} - L_{\rm sph}$
 relation by MH03 and the other lines indicate those obtained by
 correcting the $K$-band relation for A-CMR (grey solid line), T-CMR
 (dotted line), and the average $B-K$ colour (dashed line).  {\it Lower
 panel}: BHMFs calculated from iLF using the above $B$-band relations
 are indicated.}  \label{cmr}
\end{figure}

We will compare BHMFs derived from iLF using the $M_{\rm BH} - L_{\rm
sph}$ relations calculated by MH03 either in the $B$ band or $K$ band.
In order to apply the $K$-band relation to iLF ($B$ band), we need to
convert the $K$-band relation to a $B$-band relation by correcting it
for $B-K$ colour. This can be performed by using the average colour of
early-type galaxy population (e.g., Marconi et al. 2004; McLure \&
Dunlop 2004; Shankar et al. 2004). We adopt $B-K = 3.75$ estimated by
Girardi et al. (2003) for local elliptical and S0 galaxies in field and
group environments. This colour is similar to that used in Marconi et
al. (2004), where $M_Z - K$ = 4.1 for ellipticals and 3.95 for S0s
($M_Z$ is Zwicky magnitude) are used in applying the $K$-band $M_{\rm
BH} - L_{\rm sph}$ relation to the LFs from the CfA survey (Marzke et
al. 1994). These colours are based on actual measurements by Kochanek et
al. (2001) and they are converted to $B-K$ = 3.59 and 3.44,
respectively, by using the equation $B = M_Z - 0.51$ given by Aller \&
Richstone (2002).

However, this correction ignores the colour-magnitude relation (CMR). In
fact, Bell et al. (2004b) suggest that the red sequence galaxies in the
COMBO-17 survey are on the CMR which is consistent with those found for
E/S0s both in clusters (Bower, Lucey, \& Ellis 1992; Terlevich, Caldwell
\& Bower 2001) and fields (Schweizer \& Seitzer 1992). Therefore, we
attempt to take this into account. Since CMRs in $B-K$ have rarely been
investigated, we model a CMR using a population synthesis code (Kodama
\& Arimoto 1997) so as to reproduce an observed CMR in a certain set of
filters and we then derive a CMR in $B-K$ and $M_B$ using the model. In
this modelling calculation, we follow the recipe by Kodama et al.
(1998), where they modeled the CMR ($V-K$ and $M_V$) measured in the
Coma cluster (Bower et al. 1992) based on the galactic wind scenario.

It needs to be mentioned that, while a CMR is normally defined using
colours within a metric aperture, the one using total colours suits
better for this study. Total colours in more luminous (larger) galaxies
are presumed to be systematically bluer than those measured within an
aperture due to more severe effects of the colour gradients (e.g.,
Peletier et al. 1990; Tamura \& Ohta 2003); a given aperture can sample
only the reddest part of a luminous (large) galaxy, but it can include
the total light of a faint (small) galaxy. The slope of a CMR using
total colours (T-CMR hereafter) is therefore expected to be flatter than
that using colours within an aperture (A-CMR hereafter). This aperture
effect on the CMR in the Coma cluster (Bower et al. 1992) has been
investigated by Kodama et al. (1998) and the slope of the T-CMR at $M_V
\leq -20$ mag is estimated to be $\sim$ 30\% flatter. Hence we consider
a CMR with a slope flatter by 30\% than the A-CMR by Bower et al. (1992)
as a T-CMR.\footnote{In Kodama et al. (1998), $H_0 = 50$ km s$^{-1}$
Mpc$^{-1}$ is assumed and hence the slope of a T-CMR would be less flat
in our cosmology ($H_0 = 70$ km s$^{-1}$ Mpc$^{-1}$). But here we aim at
seeing how BHMFs vary by considering the CMRs and precise determination
of the slope is beyond our scope. We note that colours of ellipticals
less luminous than $M_V = -20$ mag are considered to be robust to the
aperture correction and the actual T-CMR therefore has a break at $M_V
\sim -20$ mag (see Figure 3 in Kodama et al. 1998). We ignore this and
determine the zeropoint of the T-CMR so as to reproduce the total
colours of the ellipticals with $M_V \leq -20$ mag. Although this
indicates that the T-CMR gives too red colours to less luminous
ellipticals, the impact of this on a BHMF is very small and does not
affect the following discussions.}

In Figure \ref{cmr}, the $B$-band $M_{\rm BH} - L_{\rm sph}$ relations
are shown in the upper panel. Black solid line is the $B$-band $M_{\rm
BH} - L_{\rm sph}$ relation by MH03 and the other lines indicate those
obtained by correcting the $K$-band relation for the average $B-K$
colour, A-CMR, and T-CMR. In the lower panel, BHMFs calculated from iLF
using these $B$-band relations are indicated with the same line styles
as above.
It is indicated that the $K$-band relation is not fully transformed to
the $B$-band relation by MH03 even if a CMR is taken into account, and
the BHMFs come towards the less massive end than the case where the
$B$-band relation by MH03 is applied to iLF. This may imply that, while
the tightness of the $B$-band $M_{\rm BH} - L_{\rm sph}$ relation is
nearly the same as that of the $K$-band relation, the relations are not
equivalent to each other. One possible reason for this discrepancy may
be that $B$ band luminosity is less good indicator of stellar mass due
to effects of dust extinction and/or young stellar population for some
of the galaxies used in the analysis of the $M_{\rm BH} - L_{\rm sph}$
relations.
It should be noted that the BHMF comes towards the less massive end when
the average $B-K$ colour is used than when a CMR is considered because
the adopted average colour is bluer than colours of luminous early-type
galaxies. Consequently, their $B$-band luminosities are overestimated
and therefore smaller values of $M_{\rm BH}$ are assigned to spheroids
with a certain $B$-band luminosity. For the same reason, the BHMF also
depends on choice of A-CMR or T-CMR, but the difference turns out to be
very small.

\subsubsection{Choice of $M_{\rm BH} - L_{\rm sph}$ relations from
different authors} \label{diffrel}

\begin{figure}
 \centering
 \includegraphics[width=7cm,keepaspectratio,clip]{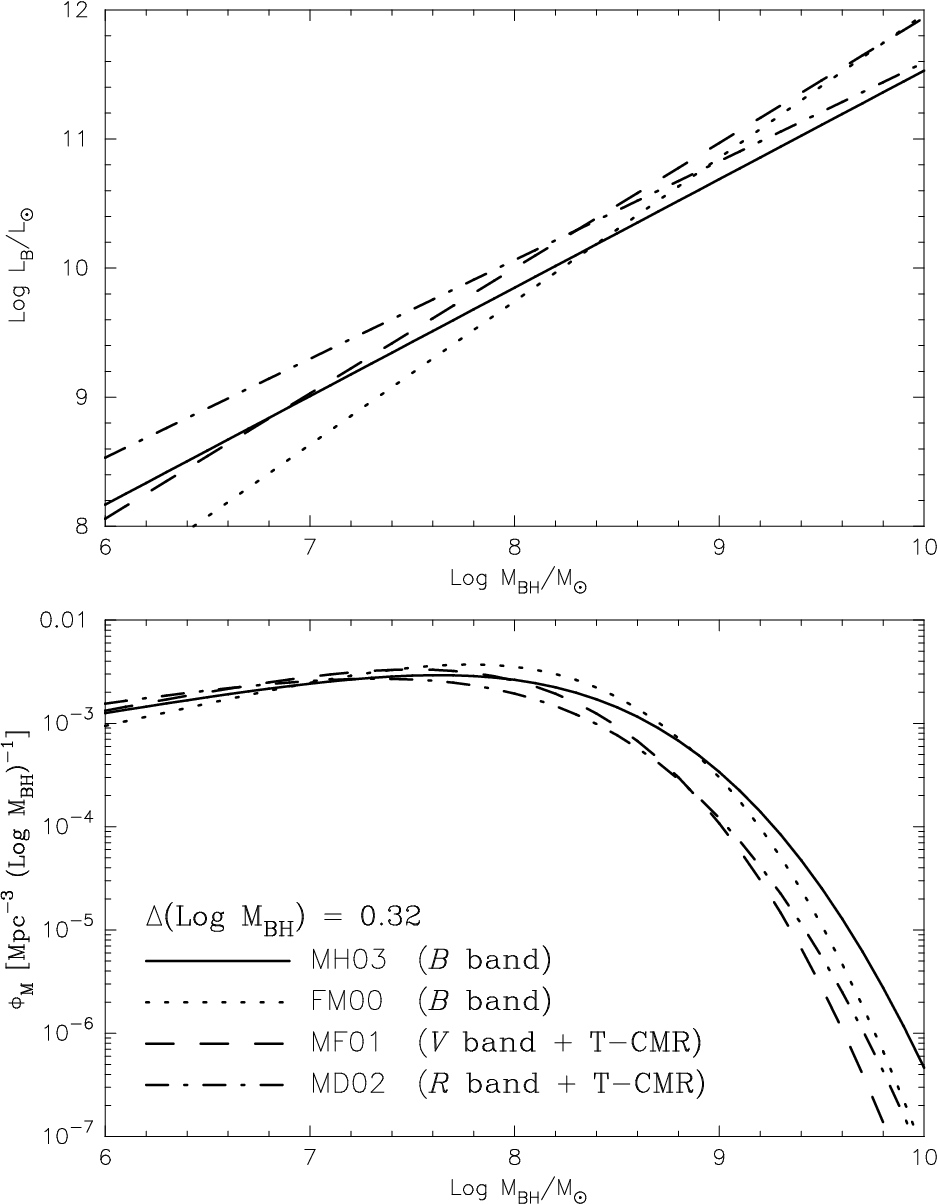}
 \caption{Same as Figure \ref{cmr}, but $M_{\rm BH} - L_{\rm sph}$
 relations by different authors and BHMFs derived using them are
 compared. {\it Upper panel}: Black solid line, grey solid line, dotted
 line, and dashed line indicates the $B$-band $M_{\rm BH} - L_{\rm sph}$
 relation by MH03, FM00, MF01, and MD02, respectively. In MD02 and MF01,
 the relation is originally derived in the $V$ and $R$ band,
 respectively, and they are converted to the $B$-band relations using
 T-CMR (see text for details). {\it Lower panel}: BHMFs derived by
 applying the above $B$-band relations to iLF are indicated. The
 intrinsic scatter of the $M_{\rm BH} - L_{\rm sph}$ relation is assumed
 to be 0.32 (MH03) in all the calculations of the BHMFs.}
 \label{authors}
\end{figure}

Next we will compare BHMFs derived using the $M_{\rm BH} - L_{\rm sph}$
relations by different authors: Ferrarese \& Merritt (2000, FM00),
Merritt \& Ferrarese (2001, MF01), McLure \& Dunlop (2002,
MD02)\footnote{Strictly speaking, since about half of the sample
consists of QSOs at $0.1 < z < 0.5$ in MD02, the $M_{\rm BH} - L_{\rm
sph}$ relation is not allowed to be used here because some evolutionary
effects may already be incorporated. In practice, however, the $M_{\rm
BH} - L_{\rm sph}$ relation derived using only the local inactive
galaxies in MD02 is the same as that from the whole sample.}, and
MH03. For FM00, we adopt the relation for their Sample A. In MF01 and
MD02, the relation is obtained in the $V$ band and $R$ band,
respectively, and they are converted to $B$-band relations using T-CMRs
modeled in the same way as explained earlier.  Also, the zeropoints of
the relations by MF01 and MD02 are shifted by the amounts due to the
differences in $H_0$ from the value we adopt.  These $B$-band relations
are shown in the upper panel of Figure \ref{authors} with the $B$-band
relation by MH03. They are applied to iLF and the BHMFs obtained are
shown in the lower panel. The intrinsic scatter of 0.32 around the
relation is assumed in all the calculations.  These BHMFs suggest that
$M_{\rm BH} - L_{\rm sph}$ relation depends on sample data set and
consequently there is a substantial variation among the BHMFs.

\subsubsection{Bulge-to-total luminosity ratio} \label{btr}

\begin{figure}
 \centering
 \includegraphics[height=7cm,keepaspectratio,angle=-90,clip]{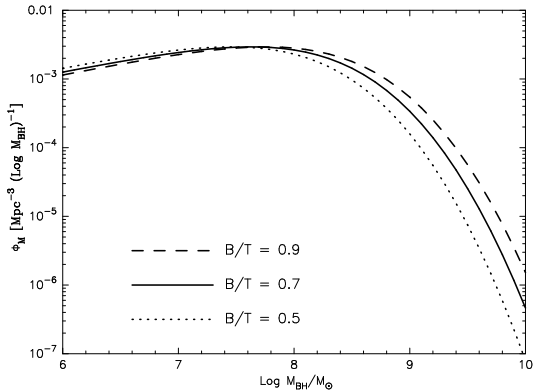}
 \caption{BHMFs for three $B/T$s are shown. These BHMFs are obtained
 from iLF and the $B$-band $M_{\rm BH} - L_{\rm sph}$ relation by MH03.}
 \label{bt}
\end{figure}

$B/T$ at intermediate redshift is only loosely constrained from
observations at the moment and we have to await for future works to see
the validity of the current assumption ($B/T = 0.7$ for all the red
sequence galaxies). It is therefore worth demonstrating BHMFs for a
range of $B/T$ to keep it in mind as uncertainty. In Figure \ref{bt},
BHMFs for $B/T =$ 0.5, 0.7, and 0.9 are compared. These BHMFs are
obtained from iLF and the $B$-band $M_{\rm BH} - L_{\rm bulge}$ relation
by MH03. This indicates that a BHMF is affected by choice of $B/T$s at
$M_{\rm BH} \geq 10^8 M_{\odot}$.

\subsubsection{Model of passive luminosity evolution} \label{pemodel}

\begin{figure}
 \centering \includegraphics[width=7cm,keepaspectratio,clip]{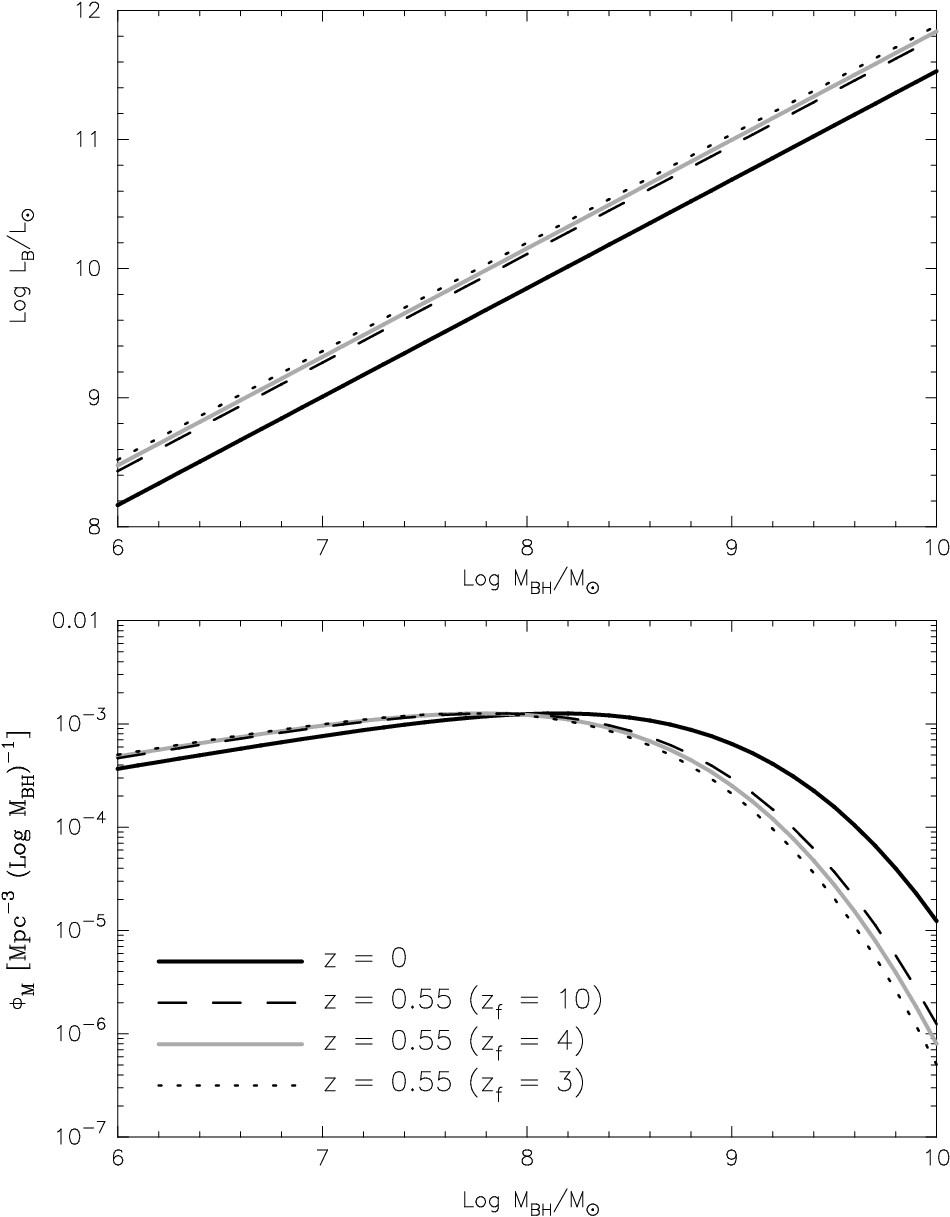}
 \caption{{\it Top panel}: The $M_{\rm BH} - L_{\rm sph}$ relations at
 $z = 0.55$ are calculated considering passive evolution of $L_{\rm
 sph}$ and are compared with the relation at $z = 0$ by MH03 (black
 solid line). We consider three models for passive luminosity evolution
 of old stellar population: $z_f = 10$ (dashed line), 4 (grey solid
 line) and 3 (dotted line). {\it Bottom panel}: BHMFs at $z = 0$ and
 0.55 derived using the above relations are indicated with the same line
 styles. Note that all the BHMFs (including that at $z = 0$) are
 calculated from the COMBO-17 LF at $z = 0.55$ (not iLF).} \label{pe05}
\end{figure}

\begin{figure}
 \centering \includegraphics[width=7cm,keepaspectratio,clip]{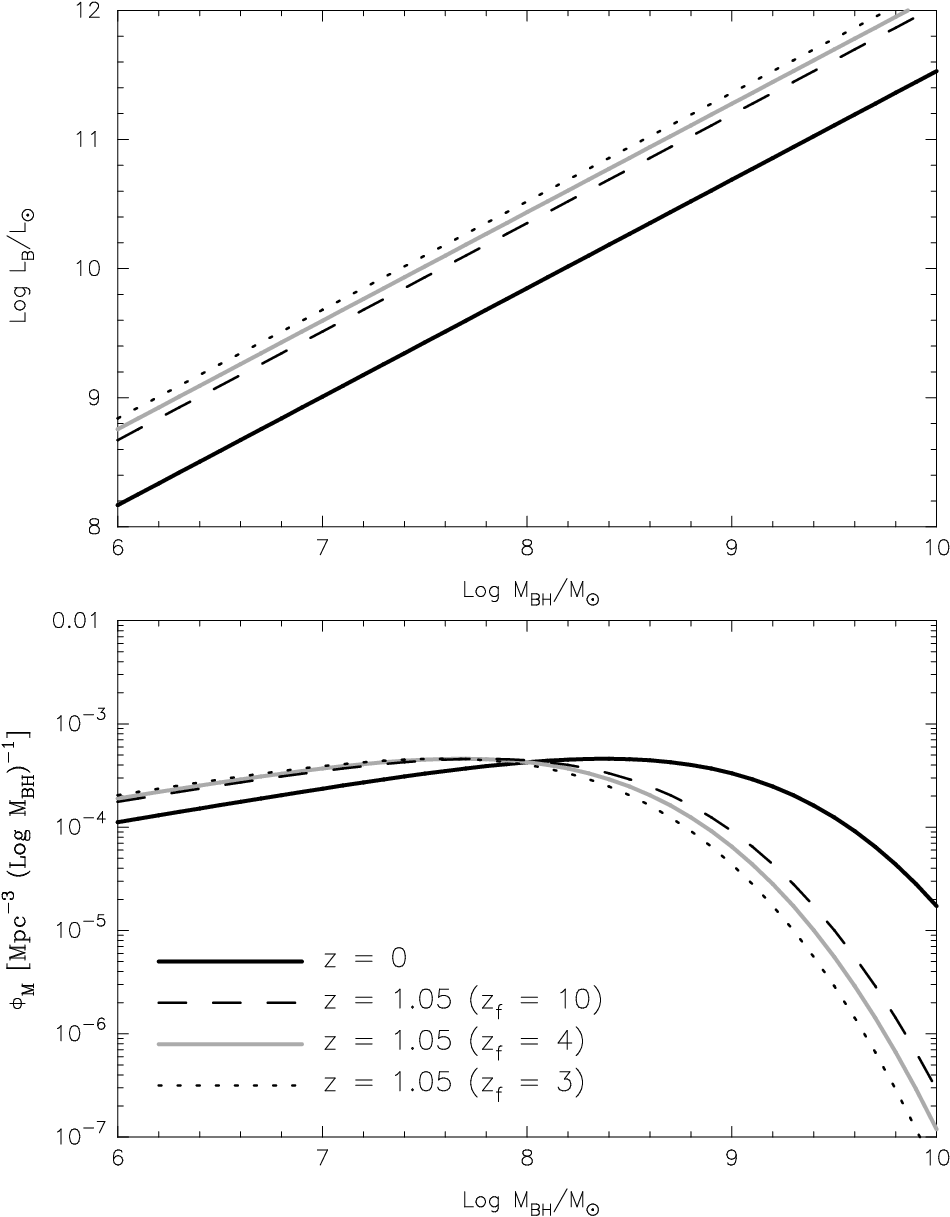}
 \caption{Same as Figure \ref{pe05}, but for $z = 1.05$.}  \label{pe10}
\end{figure}

We consider a passive luminosity evolution of an old stellar population
which formed at $z = 4$ with a starburst of which e-folding time is 1
Gyr. In this case, the luminosity evolution in the $B$ band is described
as $M_B (z) = M_B (z = 0) - Qz$ and $Q = 1.4$. The $Q$ value can be
estimated using other models of stellar populations and its dependency
on choice of parameters may need to be treated as uncertainty of BHMF at
high redshift. Since formation redshift ($z_f$) of a stellar population
is probably the most important parameter upon which the $Q$ value
largely depends, we consider two other cases: $z_f = 10$ and 3. The $Q$
value is estimated to be 1.2 and 1.6, respectively. Using these $Q$
values including 1.4 for $z_f = 4$, we derive passively evolved $B$-band
$M_{\rm BH} - L_{\rm sph}$ relations at $z = 0.55$ and apply these
relations to the COMBO-17 LF at this redshift (not iLF). The results are
displayed in Figure \ref{pe05}. We also calculate BHMFs at $z = 1.05$
using the LF and the passively evolved $M_{\rm BH} - L_{\rm sph}$
relation at this redshift and show the results in Figure \ref{pe10}.

\subsubsection{Summary of the uncertainties}\label{unsum}

\setlength{\tabcolsep}{2mm}
\begin{table}
 \centering
 \begin{tabular}{cccccccc} \hline\hline
  $\log M_{\rm BH}$ & \multicolumn{7}{c}{$\Delta \log \phi_M$} \\ \cline{2-8}
                    & (1)  & (2)  & (3)  & (4)  & (5)  & (6)  & (7)  \\ \hline
                7.5 & 0.08 & 0.02 & 0.02 & 0.11 & 0.02 & 0.02 & 0.16 \\
                8.0 & 0.04 & 0.02 & 0.07 & 0.26 & 0.09 & 0.04 & 0.20 \\
                8.5 & 0.10 & 0.01 & 0.27 & 0.37 & 0.27 & 0.15 & 0.28 \\
                9.0 & 0.37 & 0.14 & 0.57 & 0.50 & 0.53 & 0.32 & 0.41 \\
                9.5 & 0.80 & 0.45 & 0.99 & 0.96 & 0.88 & 0.54 & 0.58 \\ \hline
 \end{tabular}
  \caption{Summary of uncertainties in a BHMF at the five BH masses.
  Each number shows the possible range of logarithmic SMBH mass density
  ($\Delta \log \phi_M$) at a given $M_{\rm BH}$ caused by each source
  of uncertainty: (1) the $1 \sigma$ fitting error of the $B$-band
  $M_{\rm BH} - L_{\rm sph}$ relation (see the middle panel of Figure
  \ref{fitting}), (2) the uncertainty of intrinsic scatter ($0.3 - 0.4$)
  in the $M_{\rm BH} - L_{\rm sph}$ relation (the bottom panel of Figure
  \ref{fitting}), (3) choice of filter bands ($B$ or $K$) of $M_{\rm BH}
  - L_{\rm sph}$ relation (Figure \ref{cmr}), (4) choice of the $M_{\rm
  BH} - L_{\rm sph}$ relations by different authors (Figure
  \ref{authors}), (5) $B/T$ (0.5 $-$ 0.9; Figure \ref{bt}), (6) choice
  of passive evolution models ($z_f = 3 - 10$; $\Delta \log \phi_M$ is
  calculated at $z = 1.05$ (Figure \ref{pe10})), and (7) errors of
  characteristic luminosity and normalization in a COMBO-17 LF (the
  errors of the LF at $z = 0.65$ are considered here; see the lower
  panel of Figure \ref{relevo}). In calculating $\Delta \log \phi_M$ for
  (3), we compare a BHMF computed with the $B$-band $M_{\rm BH} - L_{\rm
  sph}$ relation obtained by correcting the $K$-band relation for the
  average $B-K$ colour (i.e., CMR is not considered) with a BHMF derived
  with the $B$-band relation by MH03. In (4), the largest and smallest
  SMBH number densities are taken at each $M_{\rm BH}$.}
  \label{errorsum}
\end{table}

Table \ref{errorsum} shows a summary of uncertainties in a BHMF
investigated earlier. In each column, the possible range of logarithmic
SMBH number density ($\Delta \log \phi_M$) caused by each source of
uncertainty is indicated at the five BH masses: $\log M_{\rm BH} =$ 7.5,
8.0, 8.5, 9.0 and 9.5. The error sources investigated are arranged in
the columns (1) $-$ (6) as follows: (1) the $1 \sigma$ fitting error of
the $B$-band $M_{\rm BH} - L_{\rm sph}$ relation (see the middle panel
of Figure \ref{fitting}), (2) the uncertainty of intrinsic scatter ($0.3
- 0.4$) in the $M_{\rm BH} - L_{\rm sph}$ relation (the lower panel of
Figure \ref{fitting}), (3) choice of filter bands ($B$ or $K$) of
$M_{\rm BH} - L_{\rm sph}$ relation (Figure \ref{cmr}), (4) choice of
the $M_{\rm BH} - L_{\rm sph}$ relations by different authors (Figure
\ref{authors}) (5) $B/T$ (0.5 $-$ 0.9; Figure \ref{bt}), and (6) choice
of passive evolution models ($z_f = 3 - 10$; $\Delta \log \phi_M$ is
calculated for BHMFs at $z = 1.05$ (Figure \ref{pe10})). In calculating
$\Delta \log \phi_M$ for (3), we compare a BHMF computed with the
$B$-band $M_{\rm BH} - L_{\rm sph}$ relation obtained by correcting the
$K$-band relation for the average $B-K$ colour (i.e., CMR is not
considered) with a BHMF derived with the $B$-band relation by MH03. In
(4), the largest and smallest SMBH number densities are taken from the
BHMFs in Figure \ref{authors}. In addition to these, $\Delta \log
\phi_M$ due to the errors of characteristic luminosity and normalization
in a COMBO-17 LF is shown for comparison in the column (7). The errors
in the LF at $z = 0.65$ are adopted to calculate $\Delta \log \phi_M$
here, but the errors in a BHMF are similar if LFs at other redshifts are
used (see the lower panel of Figure \ref{relevo}).

This table demonstrates that most of the uncertainties are negligible at
$M_{\rm BH} \leq 10^8 M_{\odot}$, while they are significant in the
range of $M_{\rm BH} \geq 10^8 M_{\odot}$. The uncertainty related to
the photometric band selection and colour correction and that related to
selection of the $M_{\rm BH} - L_{\rm sph}$ relations by different
studies are the most serious and they amount to an order of magnitude at
the massive end. Those due to the fitting error of the $B$-band $M_{\rm
BH} - L_{\rm sph}$ relation and the possible uncertainty in $B/T$ also
appear to be substantial.

\section{BHMF FROM AGN LF}\label{agnbhmf}

In this section, we investigate the cosmological evolution of BHMFs
derived from AGN LFs (AGN-BHMFs hereafter). The results will be compared
with those from COMBO-17 LFs (spheroid-BHMFs hereafter) in
\S~\ref{discussion}. By assuming that only mass accretion grows the
central SMBH at a galactic centre and galaxy mergers are not important
in its growth history, the time evolution of a BHMF $\phi_{\rm M}(M_{\rm
BH},t)$ can be described by the continuity equation:

\begin{equation}
\frac{\partial \phi_{\rm M}(M_{\rm BH}, t)}{\partial t} + \frac{\partial}{\partial M_{\rm BH}}[\phi_{\rm M}(M_{\rm BH}, t)\langle\dot{M}(M_{\rm BH}, t)\rangle] = 0,
\label{eq:continuity1}
\end{equation}

\noindent
where $\langle \dot{M}(M_{\rm BH},t) \rangle$ represents the mean mass
accretion rate at a given SMBH mass $M_{\rm BH}$ and at a cosmic time
$t$. Furthermore, if we assume a constant radiative efficiency
$\epsilon$ ($\equiv L/\dot{M}c^2$, where $L$ and $\dot{M}$ is the
bolometric luminosity and the mass accretion rate, respectively) and a
constant Eddington ratio $\lambda$ ($\equiv L/L_{\rm Edd}$, where
$L_{\rm Edd}$ is the Eddington luminosity) for all the AGNs, the second
term of the above equation can be simply related to a (bolometric)
luminosity function of AGNs. This finally reduces the continuity
equation to:

\begin{equation}
\frac{\partial \phi_M(M_{\rm BH}, t)}{\partial t} = - \frac{(1-\epsilon)\lambda^2 c^2}{\epsilon t_{\rm Edd}^2 \ln 10} \left[ \frac{\partial \psi(L,t)}{\partial L} \right]_{L = \lambda M_{\rm BH} c^2/t_{\rm Edd}},\label{eq:continuity2}
\end{equation}

\noindent
where $t_{\rm Edd}$ is the Eddington time and $\psi(L, t)$ is the AGN LF
(for details, see Marconi et al. 2004). We note that $\psi(L, t)$ is the
number of AGNs per ${\rm d} \log L$, while $\phi_M(M_{\rm BH}, t)$ is
the number of SMBHs per ${\rm d} M_{\rm BH}$. Hence, once the form of an
AGN LF is known as a function of redshift, one can integrate this
equation to obtain BHMFs at any redshifts starting from the initial
condition, either in time decreasing order (from a high redshift to
$z=0$) or the inverse (from $z=0$ to higher redshifts).

\begin{figure}
 \centering
 \includegraphics[height=7cm,keepaspectratio,angle=-90,clip]{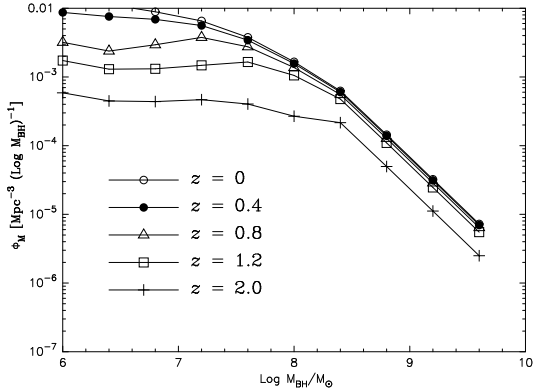}
 \caption{AGN-BHMF at a redshift of 0, 0.4, 0.8, 1.2 and 2.0 is plotted
 with open circles, solid circles, triangles, squares, and crosses,
 respectively. These AGN-BHMFs are calculated with the continuity
 equation and the HXLFs by U03 (see text for details).}\label{u03only}
\end{figure}

Following the procedure adopted by Marconi et al. (2004), here we derive
AGN-BHMFs at intermediate redshifts starting from a BHMF at $z=3$ as the
initial condition (it is assumed that all the SMBHs at $z=3$ were
shining as AGNs). In the calculation, we use the hard X-ray AGN LF
(HXLF) by Ueda et al. (2003, U03 hereafter), which is described by a
luminosity-dependent density evolution model (LDDE model; see their
\S~5.2 for details). To take into account the contribution of
``Compton-thick'' AGNs to the total mass accretion rate, we multiply a
correction factor of 1.6 independently of the AGN luminosity. The
luminosity-dependent bolometric correction described in Marconi et al.
(2004) is adopted. The Eddington ratio and the radiative efficiency are
assumed to be constant, $\lambda =1.0$ and $\epsilon =0.1$,
respectively, also based on the study by Marconi et al. (2004). The
results are shown in Figure \ref{u03only}, indicating that while the
BHMFs in the range of $M_{\rm BH} \geq 10^8 M_{\odot}$ hardly change out
to $z \sim 1$, they exhibit a clear redshift evolution at $M_{\rm BH}
\leq 10^8 M_{\odot}$. That is, almost all SMBHs with a mass larger than
$10^8 M_{\odot}$ formed at $z \ga 1$, while lighter SMBHs grow later,
suggesting a downsizing of SMBH evolution.

\begin{figure}
 \centering
 \includegraphics[width=7cm,keepaspectratio,clip]{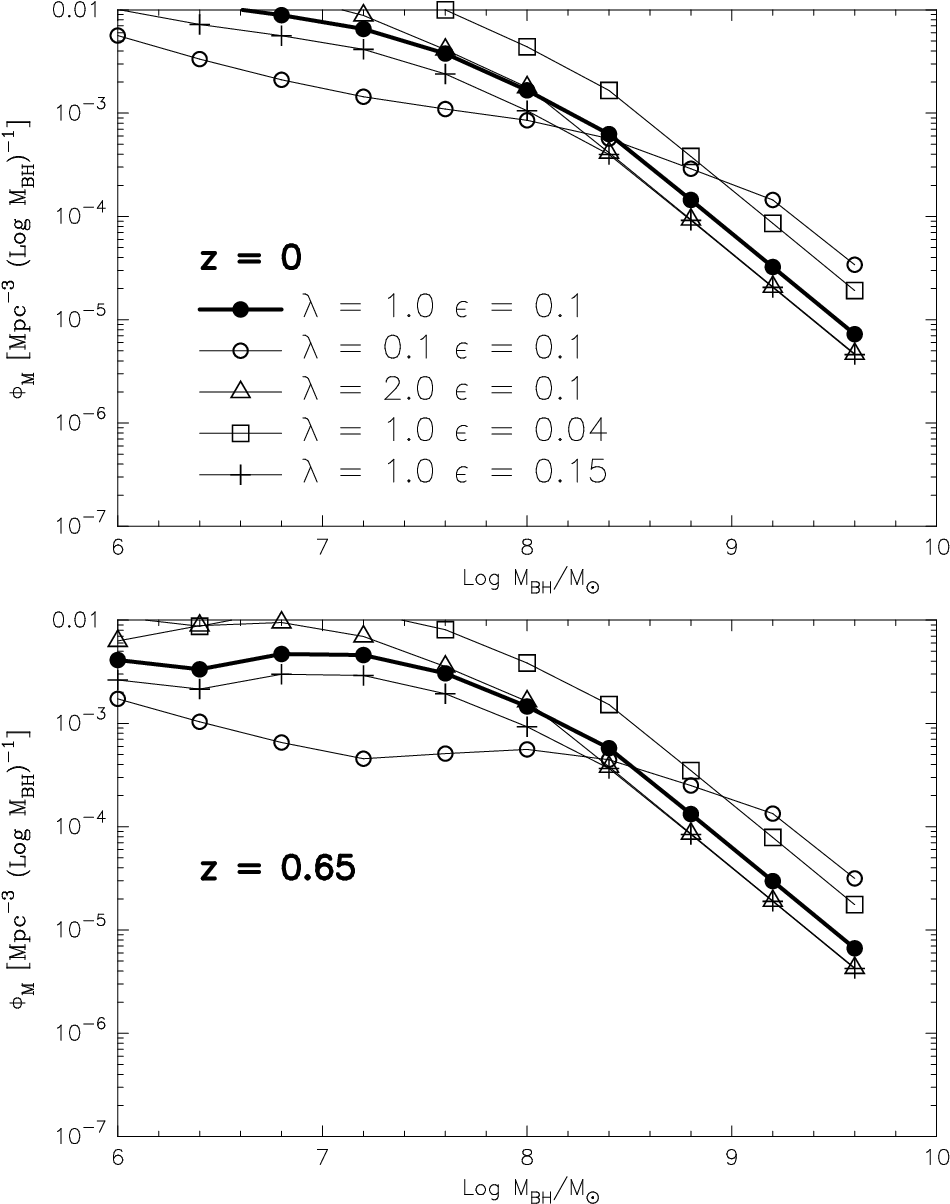}
 \caption{AGN-BHMFs at $z = 0$ (upper panel) and those at $z = 0.65$
 (lower panel) for five sets of $\lambda$ and $\epsilon$ as shown in the
 upper panel.}\label{agnbhmferr}
\end{figure}

It needs to be pointed out that there are several uncertainties in the
AGN-BHMFs thus far derived as follows:

\begin{enumerate}

 \item There are ranges of values in Eddington ratio and radiative
       efficiency which give a reasonable fit of an AGN-BHMF to the
       spheroid-BHMF at $z = 0$. The $\chi^{2}$ distribution studied by
       Marconi et al. (2004) suggests a possible range of $\lambda = 0.1
       - 2.0$ and $\epsilon = 0.04 - 0.15$ within 1 $\sigma$
       uncertainty. In Figure \ref{agnbhmferr}, we exemplify AGN-BHMFs
       at $z = 0$ and 0.65 for five sets of $\lambda$ and $\epsilon$;
       either of $\lambda$ or $\epsilon$ is fixed to the adopted value
       ($\lambda = 1.0$ or $\epsilon = 0.1$) and the largest or smallest
       value within the uncertainty is chosen for the other parameter.
       This plot indicates that, as expected from equation
       (\ref{eq:continuity2}), the normalization of an AGN-BHMF is
       altered and the AGN-BHMF is shifted along the $M_{\rm BH}$ axis
       by changing $\lambda$, while only the normalization is affected
       by changing $\epsilon$. In the following analysis, AGN-BHMFs will
       be calculated for a number of pairs of $\lambda$ and $\epsilon$
       on the 1 $\sigma$ contour provided by Marconi et al. (2004; see
       their Figure 7) and the envelope of these AGN-BHMFs that gives a
       possible range of SMBH density at a given $M_{\rm BH}$ will be
       adopted as uncertainty of AGN-BHMF.

 \item The assumption of constant Eddington ratio and radiative
       efficiency for all the AGNs is perhaps too simple. Although
       Marconi et al. (2004) claim that the local spheroid-BHMF can be
       well reproduced by the AGN-BHMF with a constant $\lambda \simeq
       1.0$ and $\epsilon \simeq 0.1$, the solution only gives a
       sufficient condition to the limited constraints at $z=0$.
       Furthermore, Heckman et al. (2004) claim that AGNs have various
       Eddington ratios and the ratios seem to depend on $M_{\rm BH}$
       Kawaguchi et al. (2004) also propose that super-Eddington
       accretion is essential for a major growth of SMBH.

 \item There is ambiguity in the continuity equation itself; if a
       merging process should be added as a source term in the equation,
       a resultant BHMF would change.

 \item Although the HXLF by U03 accounts for all the Compton-thin AGNs
       including obscured AGNs, the uncertainties in the estimate of
       Compton-thick AGNs directly affect the resulting BHMFs.

\end{enumerate}

Quantifying all of these uncertainties but (1) requires substantial
theoretical works and/or new observational data and is beyond the scope
of this paper, but one must keep them in mind.

\begin{figure}
 \centering
 \includegraphics[width=7cm,keepaspectratio,clip]{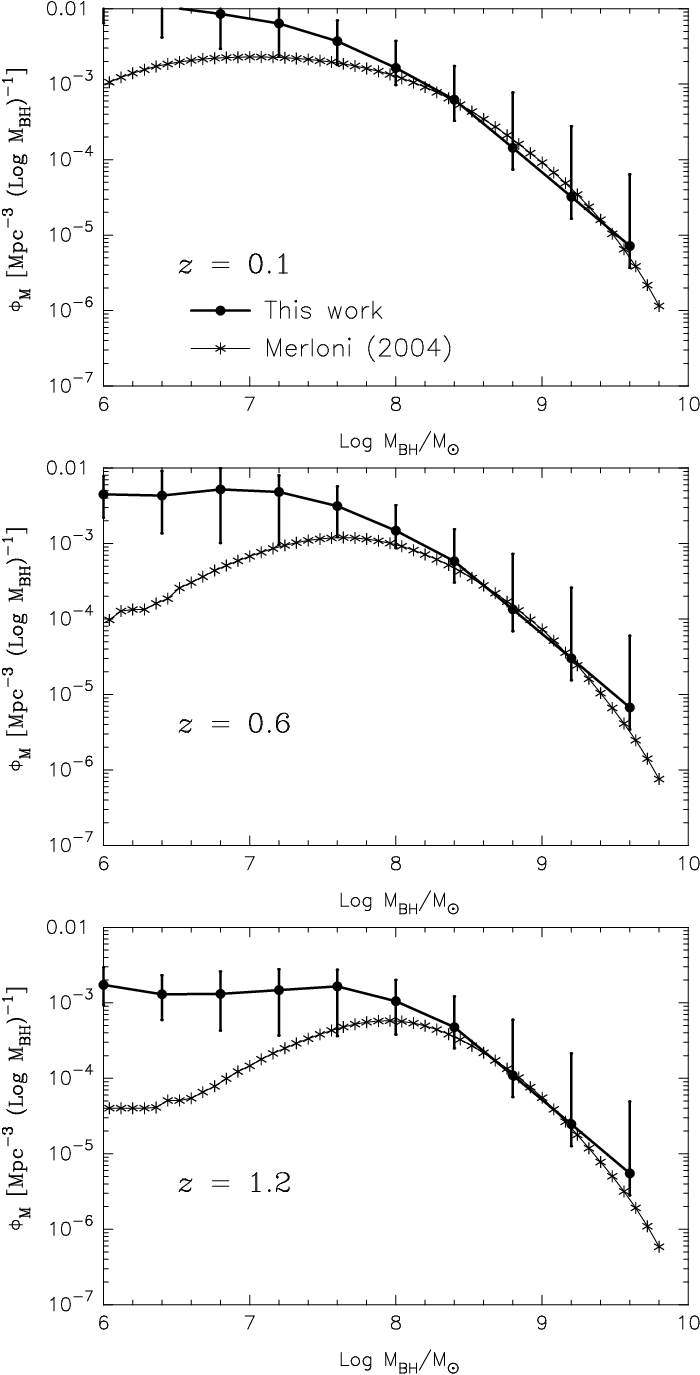}
 \caption{AGN-BHMFs at a redshift of 0.1, 0.6, and 1.2 obtained in this
 study by integrating the continuity equation starting from $z = 3$ are
 plotted with solid circles in the top, middle and bottom panel,
 respectively. The error bars represent the upper and lower bounds of
 SMBH density at a given $M_{\rm BH}$ that are calculated from AGN-BHMFs
 for the values of $\lambda$ and $\epsilon$ within the 1 $\sigma$
 uncertainty. Asterisks show AGN-BHMFs obtained by Merloni (2004).}
 \label{u03m04}
\end{figure}

It may be intriguing to compare these AGN-BHMFs with those recently
obtained by Merloni (2004), where a new method is employed to
investigate the redshift evolution. They introduce a conditional
luminosity function (CLF), which is the number of active black holes per
unit comoving volume per unit logarithm of radio ($L_{\rm R}$) and X-ray
($L_{\rm X}$) luminosity and is defined so that by integrating a CLF
over the range of $L_R$ or $L_X$, the radio LF (RLF) or X-ray LF (XLF)
of AGNs is obtained, respectively. In order to relate $L_{\rm R}$ and
$L_{\rm X}$ of AGN to $M_{\rm BH}$ without any assumptions on accretion
rate, they use an empirical relation among $L_{\rm R}$, $L_{\rm X}$, and
$M_{\rm BH}$ (fundamental plane of black hole activity; Merloni, Heinz
\& Di Matteo 2003). A CLF also needs to satisfy a constraint that the
number density of SMBHs with a certain mass is obtained by integrating a
CLF over the ranges of $L_R$ and $L_X$ which are determined by the
fundamental plane. Consequently, once a BHMF is obtained at a redshift
of $z$, a CLF can be computed using RLF, XLF, and BHMF as
constraints.\footnote{In this calculation, the fundamental plane of
black holes is assumed to be independent of redshift.} Note that RLFs
and XLFs from observations are available up to high redshifts; the RLFs
obtained by Willott et al. (2001) and the HXLF by U03 are used in
Merloni (2004). Given a functional form of accretion rate\footnote{In
Merloni (2004), $L_{\rm X}/L_{\rm Edd} = f(M, \dot{m})$ where $\dot{m}
\equiv \varepsilon_{\rm acc} \dot{M}c^2/L_{\rm Edd}$ is adopted
($\varepsilon_{\rm acc}$ is accretion efficiency).}, a mean accretion
rate can also be calculated as a function of $M_{\rm BH}$. Using a BHMF
and a mean accretion rate at $z$, a BHMF at $z + dz$ can be derived from
the continuity equation. Likewise, BHMFs at higher redshifts can
successively be calculated and hence RLF, XLF, and BHMF at $z = 0$ are
firstly needed. In Merloni (2004), the local BHMF is obtained using LFs
of galaxies with different morphologies by Marzke et al. (1994) and the
empirical relationships among spheroid luminosity, velocity dispersion,
and $M_{\rm BH}$.

Figure \ref{u03m04} shows the comparison of AGN-BHMFs at redshifts of
0.1, 0.6, and 1.2 calculated in the two different methods: (I) BHMFs
derived by integrating the continuity equation from $z=3$ to lower
redshifts using the U03 HXLF with $\lambda = 1.0$ and $\epsilon = 0.1$.
The error bars represent the upper and lower bounds of SMBH density at a
given $M_{\rm BH}$ that is calculated from AGN-BHMFs for a number of
pairs of $\lambda$ and $\epsilon$ on the 1 $\sigma$ contour provided by
Marconi et al. (2004). (II) Those obtained by Merloni (2004), which are
integrated from $z=0$ to higher redshifts using the CLF with the
fundamental-plane relation.  This comparison indicates that while the
agreement of the AGN-BHMFs is good in the massive end, the discrepancy
at $M_{\rm BH} \leq 10^{7.5} M_{\odot}$ is substantial at all
redshifts. There could be several reasons for this disagreement. One
possibility may be related to the choice of an initial spheroid-BHMF; we
note that the BHMF at $z=0.1$ in Merloni (2004) has a normalization
$\sim$ 10 times smaller at $M_{\rm BH} \sim 10^{6} M_{\odot}$ than the
local BHMF independently estimated by Marconi et al. (2004, see their
Figure 2).
Another reason could be the different estimate for the mean mass
accretion rate. In fact, we find that the second term of the equation
(\ref{eq:continuity1}) averaged between $z = 0.9$ and $z = 0.1$
calculated in method I is significantly larger than that in method II in
the range of $M_{\rm BH} \leq 10^{7.5} M_{\odot}$.
We just point out the facts in this paper and leave further discussions
for future studies. In the next section, we adopt the AGN-BHMFs
calculated by method I for comparison with the spheroid-BHMFs.

\section{DISCUSSIONS}\label{discussion}

\subsection{Comparison of Spheroid-BHMFs with AGN-BHMFs}\label{sphvsagn}

\begin{figure*}
 \centering
 \includegraphics[width=16cm,keepaspectratio,clip]{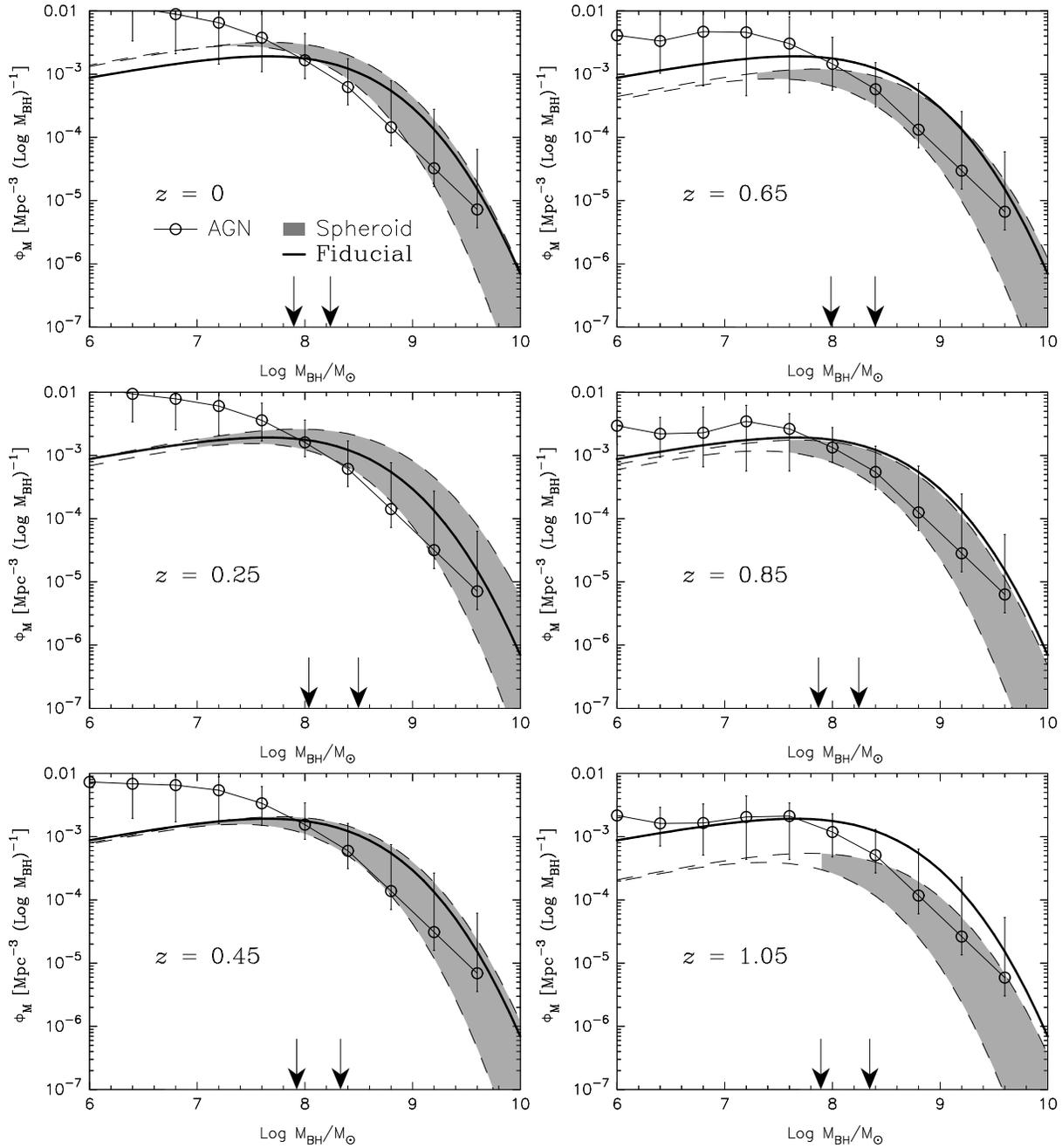}
 \caption{The spheroid-BHMFs transformed from the COMBO-17 LFs at
 redshifts of 0, 0.25, 0.45, 0.65, 0.85, and 1.05 are indicated with
 shaded regions, whose widths are determined by considering not only the
 errors of $M_{B}^{\ast}$ and $\phi^{\ast}$ but also the uncertainties
 associated with band transformation in the derivation of a BHMF (see
 text for details). The spheroid-BHMF at $z = 0$ was derived from the LF
 of the red sequence galaxies in the SDSS EDR (see Appendix of Bell et
 al. 2004b). The low mass cutoff of the shaded region corresponds to the
 lowest luminosity among the data points in the original LF. In the
 lower mass range than the cutoff, the upper- and lower-bound BHMFs are
 indicated by dashed lines. The characteristic $M_{\rm BH}$
 corresponding to $M_{B}^{\ast}$ at each redshift is indicated by arrow.
 Two arrows are shown in each panel; the one at the more massive end
 shows the characteristic $M_{\rm BH}$ of the upper-bound BHMF, and the
 other is of the lower-bound BHMF. Solid curve indicates a BHMF at $z =
 0.25$ calculated with a $B$-band $M_{\rm BH} - L_{\rm sph}$ relation
 converted from the $K$-band relation by correcting for T-CMR. This BHMF
 is displayed in all the panels as a fiducial of comparison. The
 AGN-BHMFs calculated with the continuity equation and the HXLFs by U03
 at the same redshifts as those of the spheroid-BHMFs are overplotted
 with open circles. The error bars indicate the upper and lower bounds
 of SMBH density allowing for the 1 $\sigma$ uncertainty of $\lambda$
 and $\epsilon$ as in Figure \ref{u03m04}.}  \label{passive}
\end{figure*}

In Figure \ref{passive}, spheroid-BHMFs are compared with AGN-BHMFs up
to $z \sim 1$. The spheroid-BHMFs transformed from the COMBO-17 LFs at
redshifts of 0., 0.25, 0.45, 0.65, 0.85, and 1.05 are indicated with
shaded regions (the BHMF at $z = 0$ was derived from the LF of the red
sequence galaxies in the SDSS EDR; see Appendix in Bell et al. 2004b).
The widths of the shaded regions are determined by considering not only
the errors of $M_{B}^{\ast}$ and $\phi^{\ast}$ but also the uncertainty
associated with band transformation between $B$ and $K$ of the $M_{\rm
BH} - L_{\rm sph}$ relation, which is the most significant uncertainty
among those investigated (see \S~\ref{uncertainty}). The upper bound of
the shaded region is the BHMF derived by applying the $B$-band $M_{\rm
BH} - L_{\rm sph}$ relation by MH03 to an LF with the largest values of
characteristic luminosity and normalization within the errors. To
determine the lower bound, a $B$-band $M_{\rm BH} - L_{\rm sph}$
relation calculated by correcting the $K$-band relation for the average
$B-K$ colour of early-type galaxies is applied to an LF with the
smallest values of characteristic luminosity and normalization within
the errors. The transformation of the $K$-band relation to $B$ band is
performed only at $z = 0$ and the $B$-band relations at high redshifts
are then obtained by correcting it for passive luminosity evolution in
the $B$ band. The spheroid-BHMF is depicted with shaded region down to
the mass corresponding to the lowest luminosity among the data points in
the LF. At the masses lower than this cutoff, the upper- and lower-bound
BHMFs are indicated by dashed lines. The characteristic $M_{\rm BH}$
corresponding to $L_{B}^{\ast}$ at each redshift is indicated by
arrow. Two arrows are shown in each panel; the one at the more massive
end shows the characteristic $M_{\rm BH}$ of the upper-bound BHMF, and
the other is of the lower-bound BHMF. The solid curve, which goes
through the middle of the upper- and lower-bound BHMFs at $z = 0.25$,
indicates the BHMF at $z = 0.25$\footnote{We adopt the BHMF not at $z =
0$ but at $z = 0.25$ just for consistency; the LFs at $z \geq 0.25$ are
calculated with the data from the COMBO-17 survey, while the LF at $z =
0$ is from the SDSS data.} calculated with a $B$-band $M_{\rm BH} -
L_{\rm sph}$ relation converted from the $K$-band relation by correcting
for T-CMR. This BHMF is plotted in all the panels as a fiducial of
comparison. The AGN-BHMFs calculated with the continuity equation and
the HXLFs by U03 at the same redshifts as the spheroid-BHMFs are
overplotted with open circles. The error bars indicate the upper and
lower bounds of SMBH density allowing for the 1 $\sigma$ uncertainty of
$\lambda$ and $\epsilon$.

From Figure \ref{passive}, it is suggested that at $M_{\rm BH} \geq
10^{8} M_{\odot}$, the spheroid-BHMFs\footnote{In Figure \ref{passive},
the spheroid-BHMF at $z = 0.25$ appears to exceed that at $z = 0$. This
is because of the large increase of characteristic luminosity in the
COMBO-17 LF from $z = 0$ to 0.25. This increase is significantly larger
than that predicted for passive evolution, while the rate of luminosity
evolution at $z \geq 0.25$ is fully consistent with passive evolution
out to $z \sim 1$. The origin of the large luminosity increase at the
low redshift is currently unknown.} are broadly consistent with the
AGN-BHMFs out to $z \sim 1$. This agreement between the spheroid-BHMFs
and the AGN-BHMFs appears to support that most of the SMBHs are hosted
by massive spheroids already at $z \sim 1$ and they evolve without
significant mass growth since then.
The discrepancy at $M_{\rm BH} \leq 10^{7.5} M_{\odot}$ between the
spheroid-BHMFs and the AGN-BHMFs is presumed to be due at least partly
to the fact that small bulges in late-type galaxies tend to be excluded
in selecting the red sequence galaxies and thus their contributions are
not expected to be included in the COMBO-17 LFs or the
spheroid-BHMFs. In fact, galaxy LFs are not well constrained down to
such low luminosities and future observations therefore need to be
awaited for any further discussions on the discrepancies in the light
end of BHMF.
It is interesting to point out that while the AGN-BHMFs at $M_{\rm BH}
\geq 10^8 M_{\odot}$ do not significantly evolve out to $z \sim 1$, the
spheroid-BHMFs exhibit a slight redshift evolution (see also Figure
\ref{relevo} and Figure \ref{u03only}). We note that the uncertainties
of the spheroid-BHMFs considered in Figure \ref{passive} are larger than
those in Figure \ref{relevo}, where the evolution in spheroid-BHMF may
look clearer. One possible reason for the difference in evolution
between the spheroid-BHMFs and the AGN-BHMFs is a selection effect on
the red sequence galaxy LFs; if there are more blue spheroids with
on-going star formation and/or young stellar population towards $z = 1$
then their contribution is more likely to be missed from the red
sequence galaxy LF and the spheroid-BHMF at higher redshifts. In fact,
as mentioned earlier, the BHMFs obtained from morphologically selected
early-type galaxy LFs tend to exceed those from the COMBO-17 LFs at $z
\sim 1$ (Figures \ref{im} and \ref{cross}).

\subsection{Does the correlation between $M_{\rm BH}$ and host spheroid
mass evolve with redshift?} \label{evorel}

An alternative interpretation of the possible difference in evolution
between the spheroid-BHMFs and the AGN-BHMFs may be a difference at high
redshift between the actual $M_{\rm BH} - L_{\rm sph}$ relation and our
assumption. In other words, there may be an evolution of the $M_{\rm BH}
- L_{\rm sph}$ relation other than the passive luminosity evolution of
spheroid. In order to see whether this can be the case, it is worth
examining how a BHMF can be affected by changing the $M_{\rm BH} -
L_{\rm sph}$ relation.

Here, we consider a simple case where, in the $M_{\rm BH} - L_{\rm sph}$
relation ($\log M_{\rm BH} = p \log L_{\rm sph} + q$), the coefficient
of $p$ or $q$ varies. Figures \ref{pe05} and \ref{pe10} demonstrate the
effect on BHMF of an evolution of $q$; a BHMF is shifted mostly in
parallel to the $M_{\rm BH}$ axis. The effect of changing $p$ on a BHMF
is similar, although it can be a modification in shape of BHMF rather
than a lateral shift. Therefore, changing neither $p$ nor $q$ moves the
BHMF along the axis of the number density of SMBH. On the other hand,
the redshift evolution of the actual spheroid-BHMFs seems to be
dominated by that along the number density axis, suggesting that it is
difficult to explain the slight difference in evolution between the
spheroid-BHMF and the AGN-BHMF by changing $p$ and $q$ with redshift.

\begin{figure}
 \centering
 \includegraphics[width=7cm,keepaspectratio,clip]{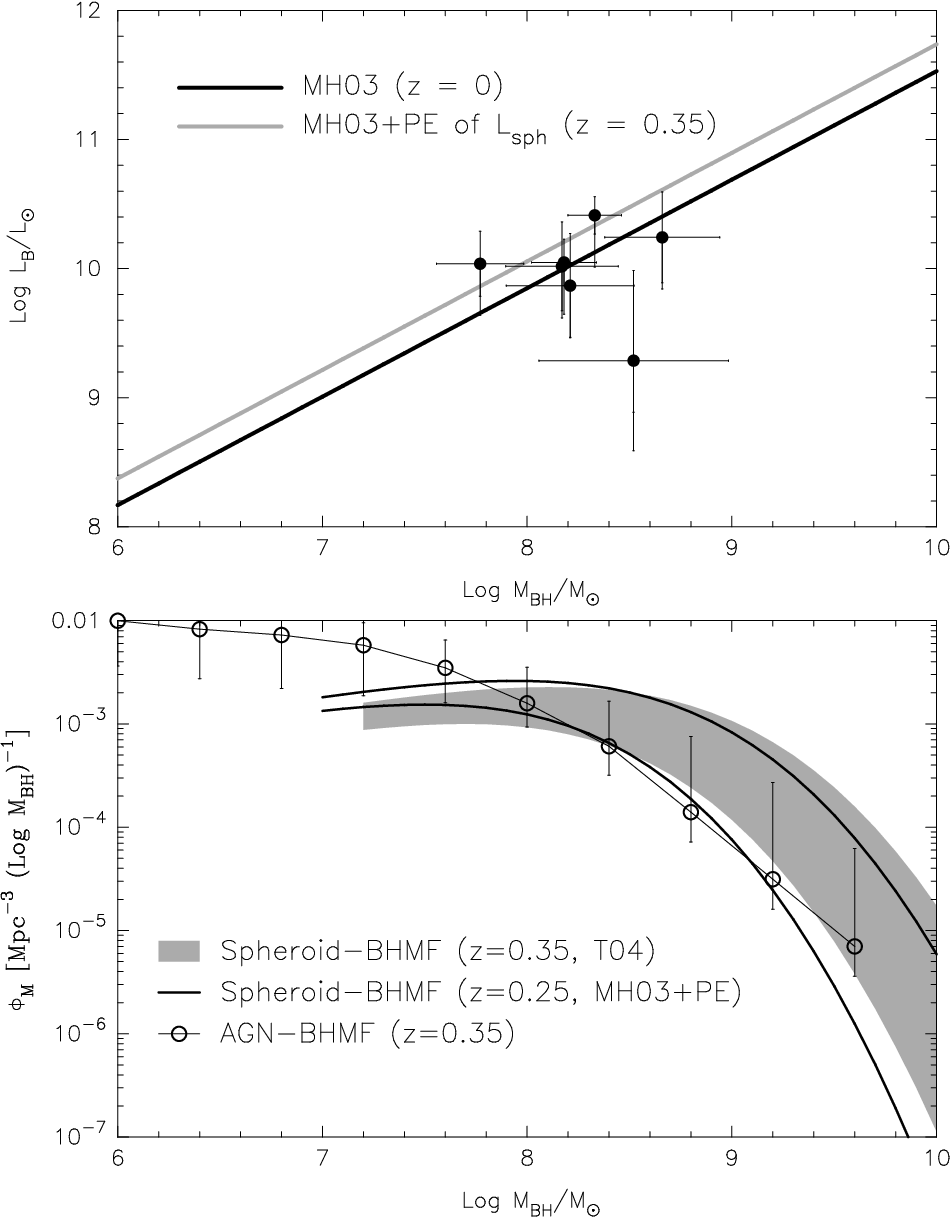}
 \caption{{\it Upper panel}: The $M_{\rm BH} - L_{\rm sph}$ relations at
 $z = 0$ (MH03; black line) and 0.35 (grey line) are shown. For the
 relation at $z = 0.35$, passive evolution of $L_{\rm sph}$ is taken
 into account. The data points indicate the galaxies observed by T04
 (see text for details). {\it Lower panel}: Shaded region indicates a
 spheroid-BHMF and its uncertainty obtained from the COMBO-17 LF at $z =
 0.35$ and the $M_{\rm BH} - L_{\rm sph}$ relation at $z = 0$, which
 seems to be followed by the T04 galaxies. Solid lines show a BHMF at $z
 = 0.25$ obtained with a $M_{\rm BH} - L_{\rm sph}$ relation considering
 a passive luminosity evolution between $z = 0$ and 0.25. The upper and
 lower bounds are determined in the same way as those in Figure
 \ref{passive}. The AGN-BHMF at $z = 0.35$ is overplotted with open
 circles. The error bars are calculated in the same way as those in
 Figure \ref{u03m04}.}  \label{treu}
\end{figure}

Recently, a possible offset from the local relationship between $M_{\rm
BH}$ and central velocity dispersion ($\sigma_0$) has been found at $z
\sim 0.37$ by investigating spectra of the central regions of galaxies
hosting type 1 AGNs (Treu, Malkan \& Blandford 2004, T04 hereafter). It
is interesting to see a spheroid-BHMF at this redshift derived with the
offset found by T04. In order to apply the relation at $z = 0.37$ to a
COMBO-17 LF at $z = 0.35$, the $M_{\rm BH} - \sigma_0$ relation needs to
be converted to a $M_{\rm BH} - L_{\rm sph}$ relation. One possible way
is to estimate spheroid luminosities of the galaxies observed by T04
from their central velocity dispersions\footnote{According to T04, the
velocity dispersions within an aperture used by T04 ($\sigma_{ap}$) are
converted to the central velocity dispersions ($\sigma_0$) as $\sigma_0
= 1.1 \sigma_{ap}$.} with the Faber-Jackson relation at $z \sim 0.4$ in
the rest-frame $B$ band (Ziegler et al. 2005) and plot them against
$M_{\rm BH}$. The results are shown with solid circles in the upper
panel of Figure \ref{treu}. Although the sample size is small and there
is a substantial scatter, the distribution of the data points suggests
that the actual $M_{\rm BH} - L_{\rm sph}$ relation followed by these
spheroids lies closer to the local relationship than that allowing for a
passive evolution of $L_{\rm sph}$ between $z = 0$ and 0.37.
In the lower panel of Figure \ref{treu}, a BHMF derived from the
COMBO-17 LF at $z = 0.35$ using the $M_{\rm BH} - L_{\rm sph}$ relation
at $z = 0$, which seems to be followed by the T04 galaxies, is indicated
with shaded region. This BHMF is compared with that at a lower redshift
($z = 0.25$) derived with a $M_{\rm BH} - L_{\rm sph}$ relation
considering a passive luminosity evolution between $z = 0$ and 0.25. The
latter BHMF is indicated with solid lines. The uncertainties of these
spheroid-BHMFs are indicated with the width of the shaded region or the
separation of the lines; the upper and lower bounds are defined in the
same way as those in Figure \ref{passive}. The AGN-BHMF at $z = 0.35$ is
overplotted with open circles. The error bars are calculated in the same
way as those in Figure \ref{u03m04}.  Although the uncertainties of the
BHMFs are large, this comparison suggests that the BHMF obtained with
the $M_{\rm BH} - L_{\rm sph}$ relation followed by the T04 galaxies is
slightly more biased to the massive end. This trend could be seen more
clearly when the BHMF is compared with that at $z = 0$. This apparently
suggests a growth of SMBH towards {\it higher} redshift, which is
unlikely in practice (see also Robertson et al.  2005). More galaxies
need to be investigated to examine the correlation between $M_{\rm BH}$
and $\sigma_e$ or $L_{\rm sph}$ at high redshift.

\subsection{Future perspective}

If much fainter end of early-type galaxy LF was determined from
observations, BHMF could be probed further down to the low mass end
($M_{\rm BH} \ll 10^8 M_{\odot}$). Since a redshift evolution of
AGN-BHMF is suggested to be fast at intermediate redshifts in this mass
range (Figure \ref{u03only}; see also Merloni 2004), it would be
interesting to look at the counterpart in spheroid-BHMFs in order to put
much stronger constraints on the co-evolution of AGN and spheroid,
particularly down-sizing effects of their evolutions. Currently, galaxy
LFs are not well constrained down to such low luminosities, especially
at cosmological distances. Even at low redshifts, the faint end slope of
an LF tends to be fixed to a certain value in fitting a Schechter
function to the data and deriving $L^{\ast}$ and $\phi^{\ast}$. Much
deeper data (e.g., $\sim$ 2 mag deeper than the COMBO-17 limit) are
essential to address the faint end of LF out to $z \sim 1$. Keeping a
survey area similar to or even wider than COMBO-17 is the key to
deriving reliable LFs and to estimate cosmic variance. Multi-band
photometry would be required to obtain photometric redshifts in good
accuracy down to the faint end. In addition, high spatial resolution
images would enable one to directly measure spheroid luminosities of
galaxies. Although it is highly expensive to take data sets satisfying
all these requirements, one promising candidate for this challenge is
the COSMOS survey: a 2 square degree field is surveyed with the HST/ACS
in the $I_{814}$ band down to 27.8 mag ($5 \sigma$) in the AB magnitude
(cf. galaxies with $R \leq 24$ mag are used to study LFs in COMBO-17)
and also with Subaru/Suprime-Cam in the
$BVr^{\prime}i^{\prime}z^{\prime}$ bands (Taniguchi et al. 2005).

\section*{ACKNOWLEDGMENTS}

We are grateful to the anonymous referee for helpful comments to improve
this paper. This research was partly supported by a Grant-in-Aid for
Scientific Research from Japan Society for the Promotion of Science
(17540216).

\label{lastpage}

\end{document}